\documentclass[twocolumn]{aastex61}

\received{2018 February 25}
\revised{2018 May 3}
\accepted{2018 June 4}
\submitjournal{ApJ}

\shorttitle{Spots, flares, accretion, and obscuration in DQ Tau}
\shortauthors{K\'osp\'al et al.}
\begin{document}

\title{Spots, flares, accretion, and obscuration in the pre-main
  sequence binary DQ Tau}

\author{\'A. K\'osp\'al} \affiliation{Konkoly Observatory, Research
  Centre for Astronomy and Earth Sciences, Hungarian Academy of
  Sciences, Konkoly-Thege Mikl\'os \'ut 15-17, 1121 Budapest, Hungary}
  \affiliation{Max Planck Institute for Astronomy, K\"onigstuhl 17,
  69117 Heidelberg, Germany} \email{kospal@konkoly.hu}
  
\author{P. \'Abrah\'am} \affiliation{Konkoly Observatory, Research
  Centre for Astronomy and Earth Sciences, Hungarian Academy of
  Sciences, Konkoly-Thege Mikl\'os \'ut 15-17, 1121 Budapest, Hungary}

\author{G. Zsidi} \affiliation{Konkoly Observatory, Research
  Centre for Astronomy and Earth Sciences, Hungarian Academy of
  Sciences, Konkoly-Thege Mikl\'os \'ut 15-17, 1121 Budapest, Hungary}

\author{K. Vida} \affiliation{Konkoly Observatory, Research
  Centre for Astronomy and Earth Sciences, Hungarian Academy of
  Sciences, Konkoly-Thege Mikl\'os \'ut 15-17, 1121 Budapest, Hungary}

\author{R. Szab\'o} \affiliation{Konkoly Observatory, Research
  Centre for Astronomy and Earth Sciences, Hungarian Academy of
  Sciences, Konkoly-Thege Mikl\'os \'ut 15-17, 1121 Budapest, Hungary}

\author{A. Mo\'or} \affiliation{Konkoly Observatory, Research
  Centre for Astronomy and Earth Sciences, Hungarian Academy of
  Sciences, Konkoly-Thege Mikl\'os \'ut 15-17, 1121 Budapest, Hungary}

\author{A. P\'al} \affiliation{Konkoly Observatory, Research
  Centre for Astronomy and Earth Sciences, Hungarian Academy of
  Sciences, Konkoly-Thege Mikl\'os \'ut 15-17, 1121 Budapest, Hungary}

%-----------------------------------------------------------------
% ABSTRACT
%-----------------------------------------------------------------

\begin{abstract}
  DQ Tau is a young low-mass spectroscopic binary, consisting of two
  almost equal-mass stars on a 15.8\,d period surrounded by a
  circumbinary disk. Here, we analyze DQ~Tau's light curves obtained
  by {\it Kepler} K2, the {\it Spitzer} Space Telescope, and ground-based
  facilities. We observed variability phenomena, including rotational
  modulation by stellar spots, brief brightening events due to stellar
  flares, long brightening events around periastron due to increased
  accretion, and short dips due to brief circumstellar
  obscuration. The rotational modulation appears as sinusoidal
  variation with a period of 3.017\,d. In our model this is caused by
  extended stellar spots 400\,K colder than the stellar effective
  temperature. During our 80-day-long monitoring we detected 40
  stellar flares with energies up to 1.2$\times$10$^{35}$\,erg and
  duration of a few hours. The flare profiles closely resemble those
  in older late-type stars, and their occurrence does not correlate
  with either the rotational or the orbital period. We observe
  elevated accretion rate up to
  5$\times$10$^{-8}\,M_{\odot}$\,yr$^{-1}$ around each periastron. Our
  {\it Spitzer} data suggests that the increased accretion luminosity
  heats up the inner part of the circumbinary disk temporarily by
  about 100\,K. We found an inner disk radius of 0.13\,au,
  significantly smaller than expected from dynamical modeling of
  circumbinary disks. Interestingly, the inner edge of the disk is in
  corotation with the binary's orbit. DQ~Tau also shows short dips of
  $<$0.1\,mag in its light curve, reminiscent of the well-known
  ``dipper phenomenon'' observed in many low-mass young stars.
\end{abstract}


%-----------------------------------------------------------------
% INTRODUCTION
%-----------------------------------------------------------------

\section{Introduction}
\label{sec:intro}

Pre-main sequence stars are intimately linked with their circumstellar
material. This relationship manifests in a variety of phenomena that
makes young stars highly variable at a wide range of wavelengths. The
photometric variability of young stars can be traced back to four main
origins: variable accretion, rotational modulation due to hot or cold
stellar spots, variable line-of-sight extinction, and stellar
flares.

According to the magnetospheric accretion model for T Tauri stars
\citep[e.g.,][]{hartmann1994}, the circumstellar disk is truncated by
the star's magnetic field and material is channeled through accretion
columns onto the star. Unsteady accretion leads to a variable
accretion rate, which may cause variability at optical wavelengths by
up to a few mangitudes on timescales as short as a few hours,
typically observed in classical T~Tauri stars
\citep[e.g.,][]{herbst1994, gullbring1994, wood1996, stassun1999,
  romanova2008}. Hot spots form at the footpoints of the accretion
columns, where shocks form due to the conversion of kinetic energy to
heat \citep[e.g.][]{dodin2015}.

Due to their strong magnetic fields (on the order of a few kG on the
stellar surface), low-mass, late-type young stars may have extensive
dark spots on their surface, which are the scaled-up versions of
sunspots. Spots on very active young stars may cover more than half of
the star's surface and are the main cause of variability in weak-line
T~Tauri stars, causing variability on a few days timescale with
amplitudes up to a few tenths of a magnitude \citep{fernandez1998,
  lehtinen2016}. While the period of variability due to dark starspots
is very stable, the exact shape of the light curve can change in a few
weeks \citep{herbst1994}.

Some low-mass pre-main sequence stars show irregular or periodic
fadings attributed to variable circumstellar extincion, called
``dipper phenomenon'' \citep[e.g.][and references
  therein]{cody2014}. In these cases, clumps of dusty material lifted
from the disk's inner edge \citep{bodman2017} or a warp in the inner
disk \citep{bouvier1999} temporarily obscure part of the starlight.

Finally, pre-main sequence stars are known to exhibit flares,
predominantly in X-rays \citep{stelzer2000,feigelson2002,favata2005},
but in some cases also in the UV and optical/white-light
\citep{rydgren1983,vrba1988,stassun2006,frasca2009}. Analogous to
solar flares, these events are probably the consequences of energy
released in a magnetic reconnection event above an active region on
the stellar surface.

While some of the listed effects causing photometric variability have
been observed in different young stars, no object has been known to
display all of them, possibly due to observational
limitations. However, the second mission of the {\it Kepler}
spacecraft (K2) provided very precise photometry for many young stars
in the Taurus star forming region in 2017. One of them, DQ~Tau, is an
extraordinary close binary system with a circumbinary disk, which
shows very complex variability patterns. With the aim of determining
the physical origin of DQ~Tau's variability and disentangle the
different effects, here we analyze the very high cadence (about 1
minute) uninterrupted 80-day-long white light monitoring of this
system obtained during the K2 mission. We complemented the K2 data
with multifilter optical photometry to determine color changes and
mid-infrared observations with the {\it Spitzer} Space Telescope to
study the response of the inner disk emission to the variable
irradiation coming from the stellar surface.

In Section \ref{sec:target} we introduce our target in detail. In
Sec.~\ref{sec:obs} we present our observations, then we describe the
immediate results on the rotational, flare, and accretion variability
in Section \ref{sec:res}. In Section \ref{sec:dis} we interpet the
periodic results using a spot model, and put into context the detected
flares by comparing them with similar flares on M-type stars. After
analyzing the energetics and morphology of the brightening of the
system near periastron and discussing the possible reasons for the
observed variable extinction, we summarize our results in Section
\ref{sec:sum}.

%-----------------------------------------------------------------
% OUR TARGET
%-----------------------------------------------------------------

\section{Our target}
\label{sec:target}

DQ~Tau is a pre-main sequence binary, consisting of two almost
identical classical T~Tauri-type stars. The system is located in the
Taurus star forming region at a distance of 140\,pc\footnote{ The
  parallax from the Gaia DR2 catalog, converted to unbiased, correct
  distances and uncertainties using \citet{bailerjones2018}, implies a
  larger distance value of $196.4 \pm 2.0$\,pc. However, since Gaia
  parallaxes of binary stars might change in later data releases, and
  since we prefer to keep our results comparable with earlier
  literature, we use 140\,pc throughout this paper.}
\citep{kenyon1994}. Based on a recent comprehensive study of
\citet{czekala2016}, the stars orbit each other in an eccentric orbit
($a$=28.96\,$R_{\odot}$, $e$=0.568), with a period of
15.80158$\pm$0.00066\,d. At periastron they approach each other within
12.5\,$R_{\odot}$ (0.06\,au), while at apastron their separation is
45.4\,$R_{\odot}$ (0.21\,au). The masses and effective temperatures of
the primary and the secondary are $M_1 = 0.63 \pm 0.13\,M_{\odot}$,
$T_1$=3700\,K, and $M_2 = 0.59 \pm 0.13\,M_{\odot}$, $T_2$=3500\,K,
respectively. The binary is surrounded by a circumbinary
protoplanetary disk, from which gas and dust are still accreting onto
the stars. ALMA observations of \citep{czekala2016} imply that the
system is seen close to pole-on ($i$=22$^{\circ}$).

DQ~Tau has been known for its quasi-periodic optical variability,
brightening by up to 0.8\,mag around periastron epochs
\citep[e.g.,][]{mathieu1997,salter2010,tofflemire2017}. One hypothesis
to explain the observed light changes is periodically modulated
accretion from the circumbinary disk onto the star. Hydrodynamical
modeling of circumbinary disks \citep[e.g.,][]{artymowicz1996}
predicts a gap in the inner part of the system cleared by the binary's
motion. These models, however, also predict a temporary stream of
accreting material accross this gap, synchronized with the orbital
motion of the stars. This phenomenon is called pulsed accretion. The
interpretation of the brightenings in DQ~Tau as accretion events is
enforced by the detection of broad and variable H$\alpha$ emission
\citep[e.g.,][]{mathieu1997}.

Periodic occurrence of millimeter flares \citep{salter2008,salter2010}
and elevated X-ray activity near periastron
\citep{getman2011,getman2016} highlighted the importance of the
binary's magnetic field. At the periastron passage, when the stars
approach each other, their magnetic fields collide, resulting in
high-energy flares. The periodically merging and separating
magnetospheres probably create a special environment for mass
accretion, which has never been theoretically or numerically
studied. This raises the possibility that the periastron brightenings
are due to a combination of magnetic and dynamic effects.

Because DQ~Tau is the prime example for pulsed accretion, it is
important to clarify the physical origin of the brightness variations.
In order to decide on the underlying processes, \citet{tofflemire2017}
performed a long-term multifilter optical monitoring of the system,
covering ten orbital periods. They detected the periastron
brightenings, and also showed hints for apastron brightness
increases. Based on the morphology of the brightness peaks and on the
energetics of the events (compared to typical flare profiles and
activity of M-type main sequence stars), they concluded that the flux
maxima close to periastron are related to temporary accretion from the
circumbinary disk onto the stars. They also detected two magnetic
flare-like events, but these were insufficient to explain the
energetic brightenings around periastron.

%-----------------------------------------------------------------
% OBSERVATIONS AND DATA REDUCTION
%-----------------------------------------------------------------

\section{Observations and data reduction}
\label{sec:obs}

DQ~Tau was suggested as a target for the \emph{Kepler} K2 Campaign 13
in our proposal (proposal ID: GO13005, PI:
\'A.~K\'osp\'al). Observations started on 2017 March 8, and finished
on 2017 May 27, providing an 80-day long uninterrupted monitoring of
the white-light brightness of DQ~Tau with a cadence of both 30\,min
(long cadence data) and 1\,min (short cadence data).

Since during the K2 mission the spacecraft operates only with two
reaction wheels, periodic thruster firings are necessary to maintain
pointing. The drift due to the solar pressure and the subsequent
correction maneuvers every few hours introduce systematic deviations
in the K2 light curves. This can be (1) due to an inadequately chosen
photometric mask, (2) because the star falls on pixels of varying
sensitivity, and (3) also because of intrapixel variations. According
to the K2 C13 Data Release
Notes\footnote{https://keplerscience.arc.nasa.gov/k2-data-release-notes.html\\\#k2-campaign-13},
the C13 pointing and roll behavior were within the limits of that seen
in other K2 campaigns. The maximum distance between the derived and
nominal positions for any target for C13 was well under the 3-pixel
limit except for three (6--18\,h long) periods with anomalous thruster
firings: one at the very beginning of the Campaign, and two during the
last 5 days of C13. These data were discarded from further
analysis. The smear correction error of Channel 74 mentioned in the
Data Release Notes does not affect the data on our target.

\begin{figure}
  \includegraphics[angle=0,width=\columnwidth]{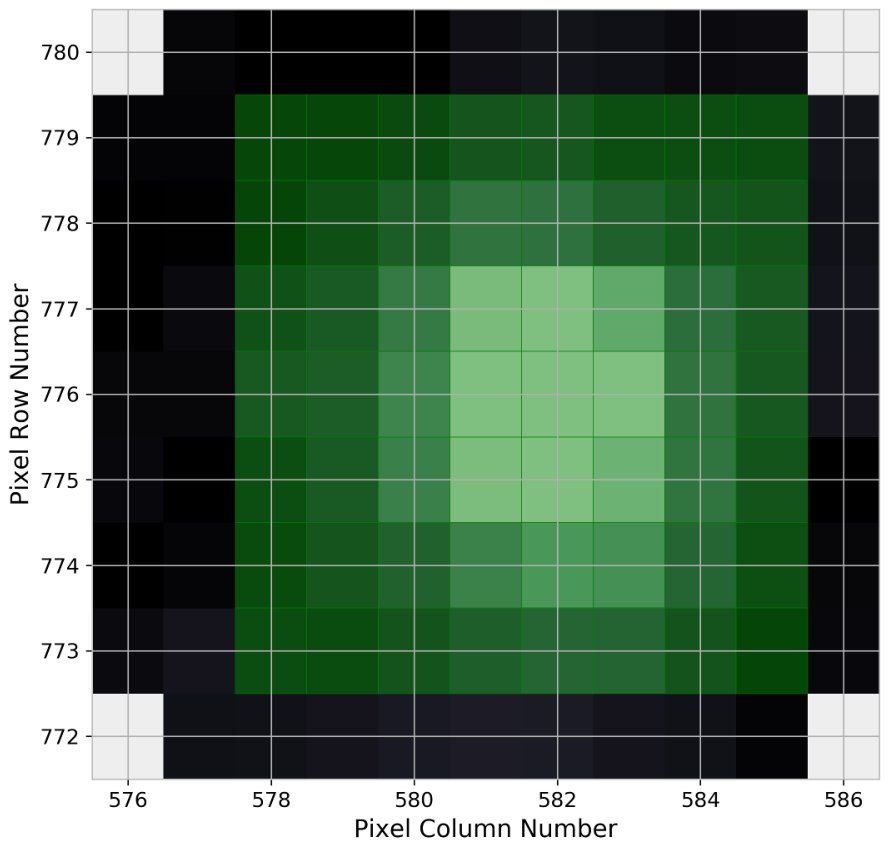}
  \caption{The large photometric aperture mask applied to DQ~Tau. The
    black pixels were downloaded from the spacecraft, the star is
    shown by the lighter shades, while the mask we applied to measure
    the flux variation of the object is highlighted in green. This set
    of pixels corresponds to the beginning of Campaign 13.
\label{fig:kepmask}}
\end{figure}

In order to extract flux variations, we used the PyKE software
package\footnote{http://pyke.keplerscience.org/} \citep{still2012}. To
sum the flux for the target, we employed a large aperture in order to
accommodate shifts due to the telescope drift. We made an aperture
file manually using the PyKE {\tt kepmask} task, then the flux was
obtained by the {\tt kepextract} task. The mask we applied for DQ~Tau
is shown in Fig.~\ref{fig:kepmask}. We followed the Extended Aperture
Method (EAP, Moln\'ar et al.~in prep.), and used a large aperture,
which provided comparable or better results (especially for large
amplitude variations) than other pipelines and took care of any
pointing systematics. We compared our extracted light curves to two
such available pipeline products, namely K2SFF (K2 Self-Flat-Fielding,
\citealt{vanderburg2014}) and EVEREST (EPIC Variability Extraction and
Removal for Exoplanet Science Targets, \citealt{luger2016}).  We found
a very good agreement (differences below 0.004\,mag) between our light
curve and the result of EVEREST. There were larger differences (up to
0.12\,mag) between our light curve and the result of K2SFF, but the
K2SFF light curve is noisier and still contains some artifacts due to
the spacecraft's maneuvers every few hours. Therefore, we concluded
that our method gives a reliable light curve for DQ~Tau and no
detrending is necessary.

\begin{figure*}
  \includegraphics[angle=90,width=\textwidth]{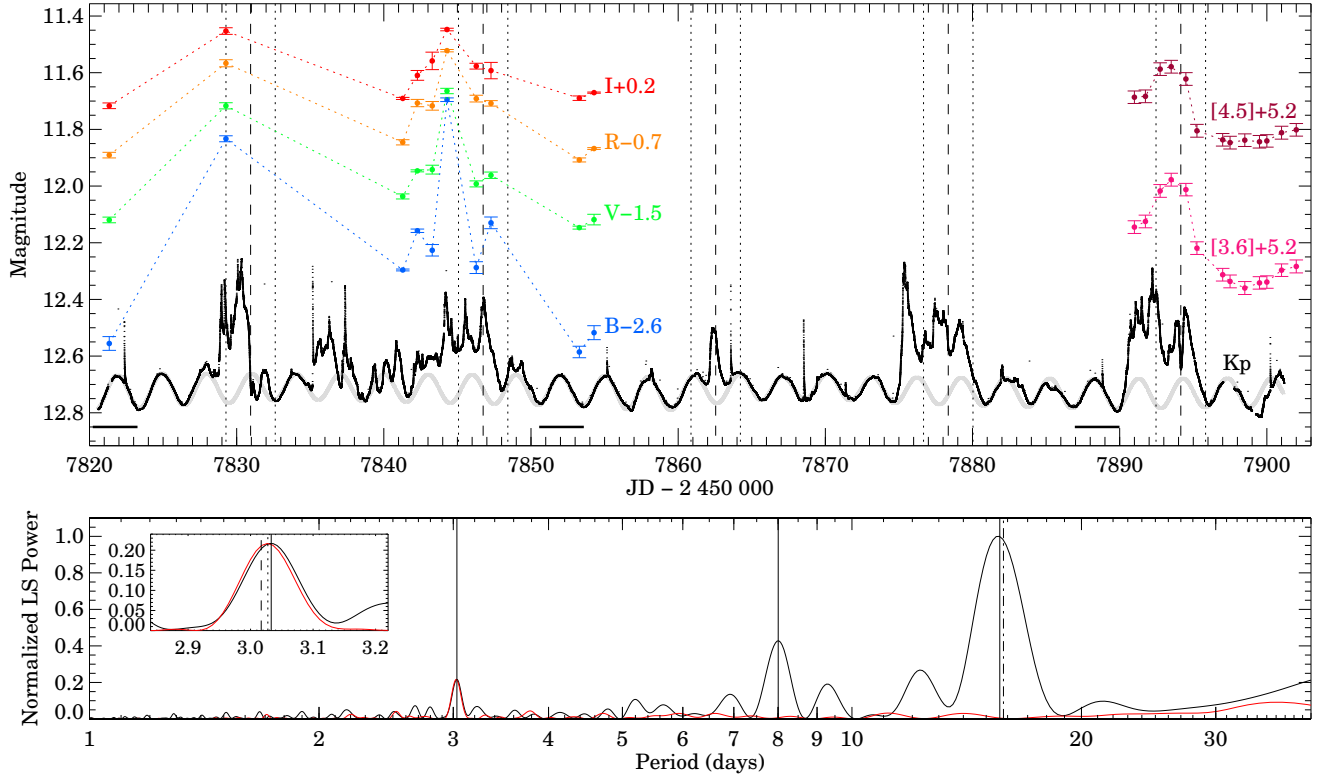}
  \caption{Top: short cadence K2 light curve of DQ~Tau (small black
    dots), along with ground-based $BV(RI)_C$ (blue, green, yellow and
    red dots) and {\it Spitzer} 3.6 and 4.5$\,\mu$m photometry (pink
    and purple dots). For clarity, the light curves were shifted along
    the y axis by the values indicated in the graph. Vertical dashed
    lines mark the periastron times using the orbital elements from
    \citet{czekala2016}. The dotted lines indicate the uncertainty by
    propagating the errors in the epoch of periastron and the orbital
    period. The thick horizonal lines mark those three cycles that are
    plotted in Fig.~\ref{fig:k2folded}. The thick gray sinusoidal
    curve indicates the rotationally modulated periodic component of
    the light curve. Bottom: Lomb--Scargle periodogram of the K2 light
    curve (with black for the full data set, with red with the dataset
    where the brightenings close to periastron were discarded). Solid
    vertical lines indicate the three most significant periods, while
    the dash-dotted line marks the spectroscopic binary's orbital
    period. The small inset shows a zoom-in around the rotational
    period. The solid line marks $P$=3.033\,d, the peak found for the
    full data set, the dotted line marks $P$=3.027, the peak found for
    the data set without the periastron brightenings, while the dashed
    line marks $P$=3.017\,d, the period needed to get no systematic
    phase shifts during the 80-day-long K2 monitoring.
\label{fig:light}}
\end{figure*}

To calibrate the K2 counts in physical units (erg\,s$^{-1}$ and $Kp$
magnitude), we first constructed the spectral energy distribution
(SED) of DQ~Tau using photometry from the VizieR database. We then
fitted a reddened stellar model from \citet{castelli2004} to the
optical data points, using $A_V = 1.5$\,mag \citep{tofflemire2017} and
an effective temperature of 3600\,K, the average of the two binary
components' temperatures (3500\,K and 3700\,K) as given by
\citet{czekala2016}. The stellar photosphere model reproduced the
observed SED very well in the 0.4--1.6$\,\mu$m wavelength range, while
the target displayed infrared excess emission at wavelengths longer
than about 1.6$\,\mu$m. We convolved the fitted photosphere model with
the high resolution response function of {\it Kepler} and integrated
it over frequency to obtain the flux of DQ Tau in the {\it Kepler}
band. Finally, we multiplied the flux by 4${\pi}d^2$ (using a distance
of $d$=140\,pc), and obtained a luminosity of
1.09$\times$10$^{32}$\,erg\,s$^{-1}$. Similarly, we used the spectrum
of Vega from \citet{bohlin2004}, convolved it with {\it Kepler}'s
response function, and integrated it over frequency to obtain Vega's
flux in the {\it Kepler} band. Assuming that Vega's magnitude is 0,
the two fluxes can be used to calculate the $Kp$ magnitude of DQ~Tau,
for which we obtained 12.722\,mag. In the following, we consider these
numbers as the average quiescent brightness of DQ~Tau and use them to
convert the K2 counts (on average 136\,000 counts in quiescence) to
luminosity or magnitude. The resulting light curve is plotted in the
top panel of Fig.~\ref{fig:light}. We also used the SED of DQ~Tau to
determine the effective wavelength of the K2 observations, obtaining
740\,nm.

For a few weeks at the beginning of the K2 monitoring, we were still
able to observe DQ~Tau from the ground in the evening twilight, using
the 60/90/180\,cm Schmidt telescope of Konkoly Observatory (Hungary),
with a 4k$\times$4k Apogee camera, and Johnson--Cousins $BV(RI)_C$
filters. Data were taken between 2017 March 8 and April 10. CCD
reduction and aperture photometry was obtained in the usual way using
an aperture radius of 4$''$ and sky annulus between 10$''$ and
15$''$. The photometric calibration of the instrumental magnitudes was
done using the APASS magnitudes \citep{apass} of 12 comparison stars
within 25$'$ of the target suggested as comparison stars for this
field by the AAVSO's finder chart
tool\footnote{https://www.aavso.org/apps/vsp/}. By checking the
ASAS-SN light curves of these stars, we confirmed that they indeed
have constant brightness \citep{shappee2014,kochanek2017}. The Johnson
$B$ and $V$ magnitudes of the comparison stars were used directly for
our calibration, while the APASS Sloan $g^{\prime}$, $r^{\prime}$, and
$i^{\prime}$ were transformed to Johnson-Cousins $R_C$ and $I_C$ with
the formulae of \citet{jordi2006}. The resulting light curves are also
plotted in the top panel of Fig.~\ref{fig:light}.

We complemented our optical data with mid-infrared photometry at 3.6
and 4.5$\,\mu$m using the {\it Spitzer} Space Telescope (proposal ID:
13159, PI: P.~\'Abrah\'am). The visibility window of {\it Spitzer}
allowed us to monitor DQ~Tau during the final 11 days of the K2
campaign, between 2017 May 17 and May 28, with an average cadence of
20\,hours. We used the IRAC instrument in full-array mode with
exposure times of 0.4\,s per frame. We used an aperture of 3 pixels
(3$\farcs$6), sky annulus between 3 and 7 pixels
(3$\farcs$6--8$\farcs$4), and aperture correction factors of 1.125 and
1.120 at 3.6 and 4.5$\,\mu$m, respectively (IRAC Instrument
Handbook\footnote{http://irsa.ipac.caltech.edu/data/SPITZER/docs/irac/\\iracinstrumenthandbook/27/}). We
performed the photometry on the corrected basic calibrated data (CBCD)
frames produced by the S19.12.0 pipeline at the Spitzer Science
Center. To derive final photometry and uncertainties, we computed the
average and the rms of the flux densities extracted from the five CBCD
images corresponding to the individual dither steps. These light
curves are also displayed in the top panel of Fig.~\ref{fig:light}.

%-----------------------------------------------------------------
% RESULTS
%-----------------------------------------------------------------

\section{Results}
\label{sec:res}

\subsection{The K2 light curve}

The top panel of Fig.~\ref{fig:light} displays the short cadence light
curve of DQ~Tau. With vertical dashed lines we overplotted the
periastron times using the orbital elements from
\citet{czekala2016}. The dotted lines indicate the uncertainty by
propagating the errors in the epoch of periastron and the orbital
period. The 80-day-long K2 light curve covers 5 orbital cycles. As
expected, there are brightening events of 0.2--0.4\,mag coinciding
with each periastron, with various durations and light curve
morphologies. There was one cycle when a similar brightening event
could be observed around the apastron. For the remaining time, the
variability is dominated by a sinusoidal pattern with slightly varying
phase and amplitude, and brief flare-like brightenings.

We calculated the Lomb-Scargle periodogram of the K2 light curve,
which is plotted with a black curve in the bottom panel of
Fig.~\ref{fig:light}. The normalized periodograms of the short and
long cadence data gave identical results. The three most significant
periods, all with False Alarm Probability below $10^{-8}$, are 15.627
$\pm$ 1.075\,d, 8.003 $\pm$ 0.262\,d, and 3.033 $\pm$ 0.044\,d. The
first one is consistent within the uncertainties with the orbital
period of 15.80158 $\pm$ 0.00066\,d of the spectroscopic binary
\citep{czekala2016}, as expected because of the brightenings around
the periastrons. The second one is half of the first period, and is a
consequence of brightenings both at periastron and (sometimes) at
apastron as well. The third and shortest period is consistent with the
rotational period of the stars reported previously by
\citet{basri1997}, indicating the presence of starspots on either or
both stars leading to rotational modulation in the light curve.

\subsection{Sinusoidal variations in the light curve}

We found an approximately 3\,d periodicity in the K2 light
curve. Remarkably, using $v{\sin}i$ of 10\,km\,s$^{-1}$, stellar
radius of 1.6\,$R_{\odot}$, and inclination of 23$^{\circ}$ from
\citet{mathieu1997}, the derived rotational period is also 3\,d. We
note, however, that while more recent radial velocity studies give
similar $v{\sin}i$ values (\citealt{nguyen2012} gave 14.7 and
11.3\,km\,s$^{-1}$ for the primary and secondary, respectively, while
\citealt{czekala2016} gave 14 and 11\,km\,s$^{-1}$), the smaller
stellar radii (1.05 and 1.00\,R$_{\odot}$ calculated from the stellar
luminosity and temperature by \citealt{tofflemire2017}) mean that
these numbers would give much shorter rotational periods, on the order
of 1.4--1.7\,d. Nevertheless, the K2 light curve clearly suggests that
either both stars have a rotational period of about 3\,d, or (if not
both stars are spotted) the spotted component has a rotational period
of 3\,d.

\begin{figure}
  \includegraphics[angle=0,width=\columnwidth]{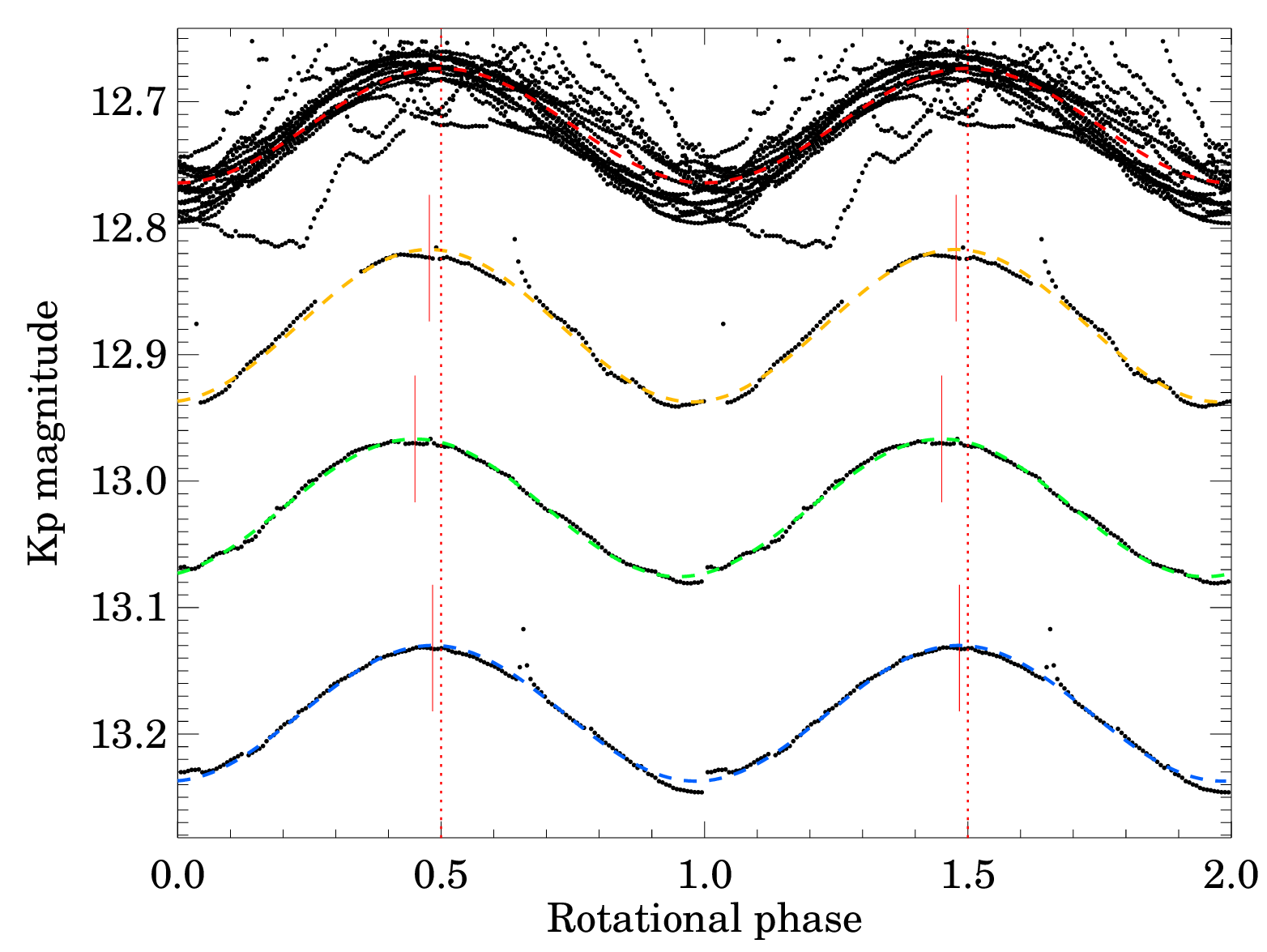}
  \caption{Phase-folded long cadence K2 light curve of DQ~Tau using
    $P$=3.017\,d, together with a sine function fitted for the full
    data set (red dashed curve). The data points and sine fits below
    display three single cycles, between JD$-$2\,450\,000 =
    7820.23--7823.26, 7850.56--7853.59, and 7886.95--7889.99. For
    clarity, these have been shifted along the y axis. The
    corresponding time intervals are marked in Fig.~\ref{fig:light}
    with thick horizontal lines. The sine fits to the single cycles
    (dashed yellow, green, and blue curves) clearly show that they
    give slightly different phases, but with this period, there is no
    systematic phase change over the 80 days of observations. To guide
    the eye, the vertical red lines mark the maxima of the sine
    functions.
\label{fig:k2folded}}
\end{figure}

In order to be able to precisely study the rotational modulation in
the K2 light curve, we needed to discard the periastron-related
brightenings and any other irregular flux variations. For this
purpose, we phase-folded our K2 light curve with the 15.80158\,d
orbital period, and found that the periastron-related activity is
mostly confined to phases between 0.75--1.1 (the periastron being at
phase 0), in accordance with the findings of
\citet{tofflemire2017}. We discarded these points, as well as those
brighter than 12.65\,mag. We recalculated the periodogram for the
remaining data points, which is plotted with a red curve in the bottom
panel of Fig.~\ref{fig:light}. The new periodogram has the strongest
peak at the stellar rotational period, but with a small shift compared
to the full data set: 3.027 $\pm$ 0.041\,d. We phase-folded the light
curve with this period, and found that the folded curve can be very
well fitted with a sine curve. However, the data points showed a large
scatter around the fitted sine curve, not explained by the photometric
uncertainty, which is on the order of 10$^{-4}$\,mag. Motivated by
this, we fitted sine functions separately to each rotational cycle, in
order to look for systematic phase shifts or amplitude
changes. Indeed, we found a gradual phase change from the first cycle
to the last, which suggested that the period determined from the
periodogram is still slightly biased. We determined the period which
resulted in no systematic phase shifts (although some phase difference
between the individual cycles still remained), and obtained
$P$=3.017$\pm$0.004\,d. Fig.~\ref{fig:k2folded} shows the K2 light
curve folded with this period and the fitted sine function, which has
an average magnitude and peak-to-peak amplitude of
${<}Kp{>}=12.7189\pm0.0001$\,mag and
${\Delta}Kp=0.0905\pm0.0001$\,mag, respectively.

%We obtained a phase shift of 0.73 compared to Tofflemire's data,
%which corresponds to a difference of about 97$^{\circ}$ (so, like a
%quarter rotation). This means that the spots were located at different
%longitude during the K2 mission compared to Tofflemire's
%observations.

\subsection{Flare detection and characteristics}
\label{sec:flares}

\begin{figure*}
  \includegraphics[angle=90,width=\textwidth]{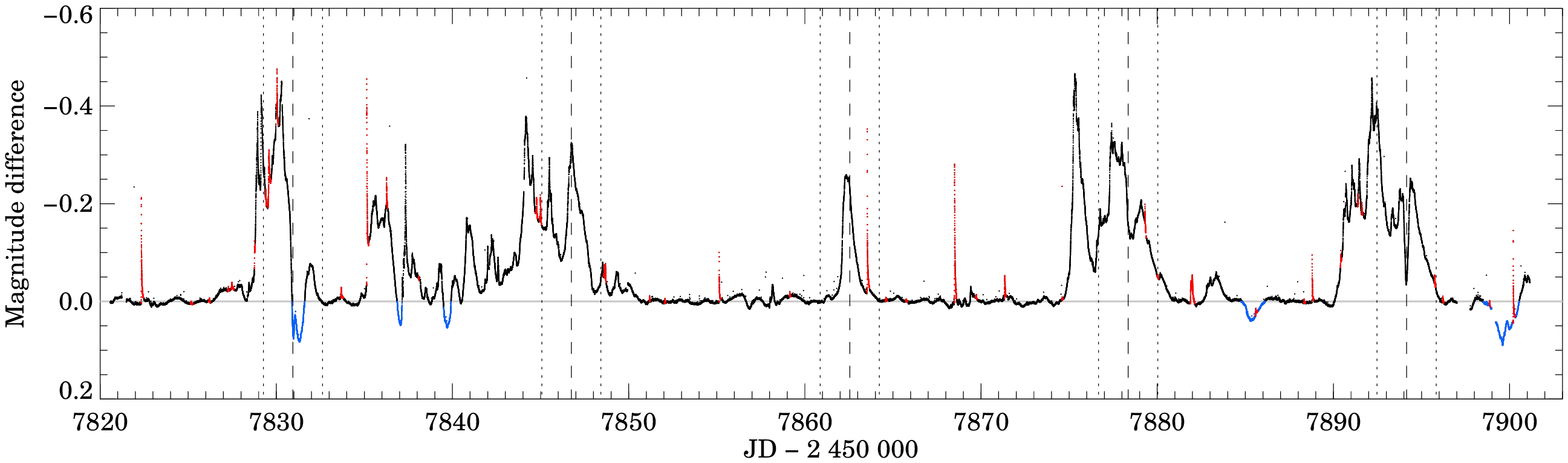}
  \caption{K2 light curve of DQ~Tau after removing the sinusoidal
    variation due to rotational modulation. Points highlighted with
    red color indicate the identified flares (Section
    \ref{sec:flares}), while the blue points mark possible ``dipper''
    events (Section \ref{sec:dipper}). As in Fig.~\ref{fig:light},
    dashed lines mark the periastrons and dotted lines indicate their
    uncertainties.
\label{fig:corrected}}
\end{figure*}

As previously described, we determined the exact phase, amplitude, and
magnitude shift of each 3.017\,d cycle. Interpolating linearly between
these numbers, we calculated the phase, amplitude, and magnitude shift
for each point in time during the K2 observations, and constructed a
sinusoidal wave with a continuously changing phase, amplitude, and
magnitude shift. This wave is plotted in Fig.~\ref{fig:light} with a
thick gray curve. Then we subtracted this sinusoidal wave from the
original light curve, thereby creating a data set where the rotational
modulation is corrected for. We used the resulting light curve,
plotted in Fig.~\ref{fig:corrected}, to look for short (typically a
few hours long) flare-like brightening events. We searched for flares
by visually inspecting the light curve and its second derivative. We
looked for events similar to a single classical flare (see, e.g.,
\citealt{gershberg1972}), which ideally consists of a fast rise phase
and an exponential decay phase, as described by \citet{hawley2014} and
\citet{davenport2014}, or complex flares consisting of multiple
eruptions. We identified 40 such events, with the caveat that we might
have missed some very faint flares that were difficult to discern from
the noise. These are highlighted with red color in
Fig.~\ref{fig:corrected}.

For each flare, we fitted a second or third order polynomial baseline
to the points preceding and following the flare, and subtracted
it. The resulting flare light curves (normalized to their peak values
for better comparison) are plotted in Figs.~\ref{fig:flares1} and
\ref{fig:flares2}. In order to establish whether these flares are
similar to the classical flares typical in M-type stars, we fitted our
data points with the flare template constructed by
\citet{davenport2014}, see their Fig.~4 and Eqn.~1 and 4. We fitted
the flare templates using the Levenberg-Marquardt least-squares
minimization procedure coded in {\tt IDL} (Interactive Data Language)
by \citet{markwardt2009}. By looking at the fitted flare templates
plotted in red over our Figs.~\ref{fig:flares1} and \ref{fig:flares2},
it is evident that many of the observed flares are well fitted by the
template, but there are a few exceptions. In some cases, the rise
phase is well fitted, while the decay phase is not exactly: the
beginning of the decay phase follows the template shape, but then
there is often an excess first and a deficiency later (e.g., flares
\#6 and \#12). In other cases, especially visible for the strongest
flares, the beginning of the decay phase is steeper, while later it
becomes shallower than the template (e.g., flares \#1, \#21, or
\#40). This suggests that the double exponential with the particular
exponents \citet{davenport2014} used is not always a good
representative of DQ~Tau's flares. In some cases, the flares had
distinct triangular shapes (e.g., flare \#7), which are very different
from the classical flare template and cannot be well fitted with
it. In some cases, we could fit the light curves with the sum of
several flare templates, indicating that DQ~Tau displayed some complex
flares. Out of the 40 flares, 25 were single classical flares, while
15 were complex flares.

\subsection{Periastron brightenings}
\label{sec:periastronbrightenings}

\begin{figure*}
  \includegraphics[angle=90,width=\textwidth]{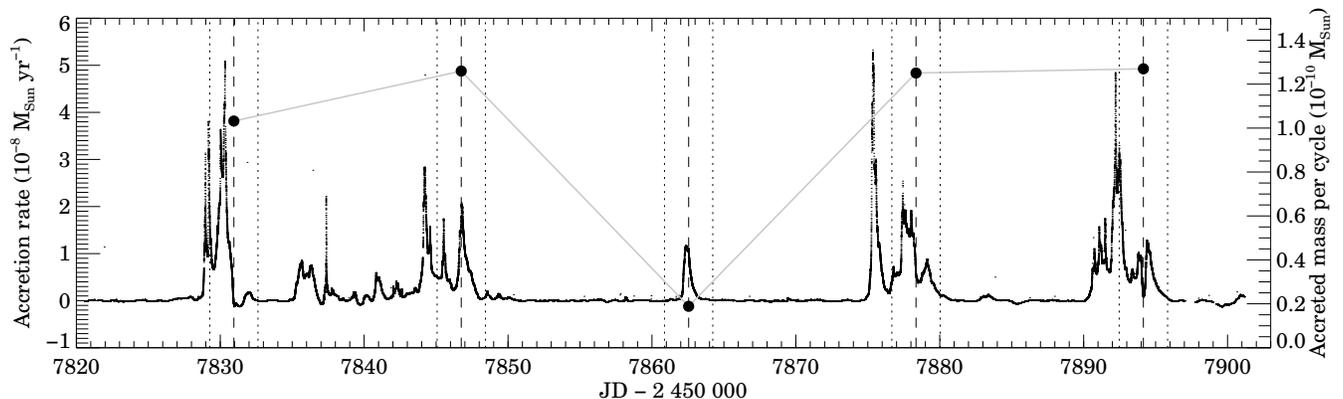}
  \caption{Accretion rate as a function of time (black curve) and mass
    accreted per orbital cycle (large black dots connected by gray
    lines) in the DQ~Tau system. As in Fig.~\ref{fig:light}, dashed
    lines mark the periastrons and dotted lines indicate their
    uncertainties.
\label{fig:acc}}
\end{figure*}

In order to construct a light curve which is devoid not only of
rotational modulation but also of flares, as a next step we subtracted
the flare templates fitted in the previous subsection from the light
curve. The remaining variability mostly consists of complex
brightening events clustered around the periastrons. It is very
unlikely that these events are stellar flares, because they last for
several days (even the shorter brightening at JD$-$2\,450\,000 =
7862.4 lasted for about 20 hours), while typical flares are shorter
than a few hours. The light curve morphology is also distinctly
different from the short brightening and exponential fading expected
for flares. We tried to fit the periastron events with the sum of
multiple flare templates, but we could not obtain adequate
fits. Another argument againts the periastron brightenings being due
to complex flares is the different color. Our multi-filter optical
monitoring covered two periastron events (Fig.~\ref{fig:light}). While
the brightening amplitudes decrease from the $B$ to $I$ band
(${\Delta}R=0.44\times{\Delta}B$, thus the object becomes bluer when
brighter, the observed colors are still redder than those of typical
stellar flares (${\Delta}R=0.17\times{\Delta}B$,
\citealt{davenport2012}). Based on these arguments, in agreement with
\citet{tofflemire2017}, we can conclude that the variability remaining
after removing the rotational modulation and the stellar flares are
most probably due to variable accretion onto the stars.

\citet{tofflemire2017} used their $U$-band light curve as a proxy for
the mass accretion rate. They first calculated the $U$-band excess
luminosity above the stellar photosphere, then used a correlation
between $L_U \rm excess$ and $L_{\rm acc}$ from \citet{gullbring1998}
to estimate the total accretion luminosity:
$$
\log(L_{\rm acc}/L_{\odot}) = 1.09 \log(L_{U \rm excess} / L_{\odot}) + 0.98,
$$
from which the accretion rate can be calculated as follows:
$$
\dot{M} = \frac{L_{\rm acc}R_*}{GM_*}\left(1-\frac{R_*}{R_{\rm in}}^{-1}\right).
$$
In order to construct the best possible time-resolution monitoring of
the accretion rate in the DQ~Tau system, we attempted to use the K2
light curve as a proxy for the accretion. First we confirmed that the
colors of the periastron events according to our multi-filter
observations and according to \citet{tofflemire2017}'s data are
consistent in the common filters ($BVR$). Then we plotted the
variability amplitudes measured by \citet{tofflemire2017} as a
function of wavelength using the effective wavelengths of the $UBVR$
filters. Then, using 740\,nm as the effective wavelength of the K2
observations for DQ~Tau, we extrapolated how much larger the
variability is in the $U$ band compared to K2. We found that the
$U$-band variability is 6.4 times that of the K2-band in magnitude
scale. Using this information, we could convert our K2 light curve
(after subtracting the sinusoidal variation and flares) to a $U$-band
light curve, then calculated the accretion rate using the same
formulae and stellar parameters as \citet{tofflemire2017}. Following
their example, we also calculated the total accreted material in each
orbital cycle (from phase 0.7 to phase 1.3). Our results can be seen in
Fig.~\ref{fig:acc}.

\begin{figure}
  \includegraphics[angle=0,width=\columnwidth]{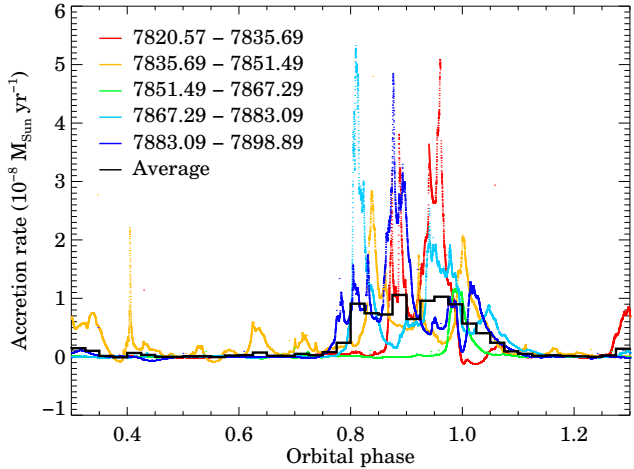}
  \caption{Accretion rate in the DQ~Tau system, phase-folded with the
    orbital period of 15.80158\,d. Different colors mark the five
    consecutive cycles that K2 covered. The corresponding
    JD$-$2\,450\,000 values are displayed in the upper left
    corner. The black histogram shows the average accretion rate along
    the orbit in 0.025 wide phase bins. Phase 1.0 corresponds to the
    periastron.
\label{fig:acc2}}
\end{figure}

As it was already evident from the light curve in
Fig.~\ref{fig:light}, the accretion rate curve in Fig.~\ref{fig:acc}
confirms that the brightenings caused by the increased accretion rate
are happening preferentially near periastron. The phase-folded
accretion rate curve and its average in Fig.~\ref{fig:acc2} shows that
significant accretion mostly happens between phases 0.75 and 1.1, as
already noted by \citet{tofflemire2017}. The strongest peaks reach
5$\times$10$^{-8}\,M_{\odot}$\,yr$^{-1}$, when the accretion
luminosity goes up to 0.8$\,L_{\odot}$. We observed increased
accretion at each periastron covered by the K2 monitoring. The total
mass accreted in each orbital cycle is typically about
1.2$\times$10$^{-10}\,M_{\odot}$, except for a rather weak event at
JD$-$2\,450\,000 = 7862.4. The unprecedented precision and high
cadence of the K2 observations reveal fine details in the accretion
rate changes, like a typical three-peaked shape, which is evident not
only in some individual events but also in their average plotted by a
black histogram in Fig.~\ref{fig:acc2}. We note that, to our best
knowledge, such an accretion rate curve shape is not predicted by any
numerical simulations of accretion in a binary system. While two peaks
could be explained by a time difference in the maximal accretion onto
the two stars, we have no explanation for three peaks.

A more detailed inspection of the periastron events hints at another
periodic phenomenon in the accretion rate. For example, in the last
monitored periastron event, the accretion rate displays several narrow
maxima which seem to follow each other at multiples of 8.8\,h
(Fig.~\ref{fig:acc}). Similar phenomenon although with different
period might be present at the other periastron events as well. This
suggests that the accretion process is not smooth, but the star
probably accretes clumpy material.

%-----------------------------------------------------------------
% DISCUSSION
%-----------------------------------------------------------------

\section{Discussion}
\label{sec:dis}

\subsection{Spots on the stars}
\label{sec:spotmodel}

\begin{figure}
\begin{center}
  \includegraphics[angle=0,width=\columnwidth]{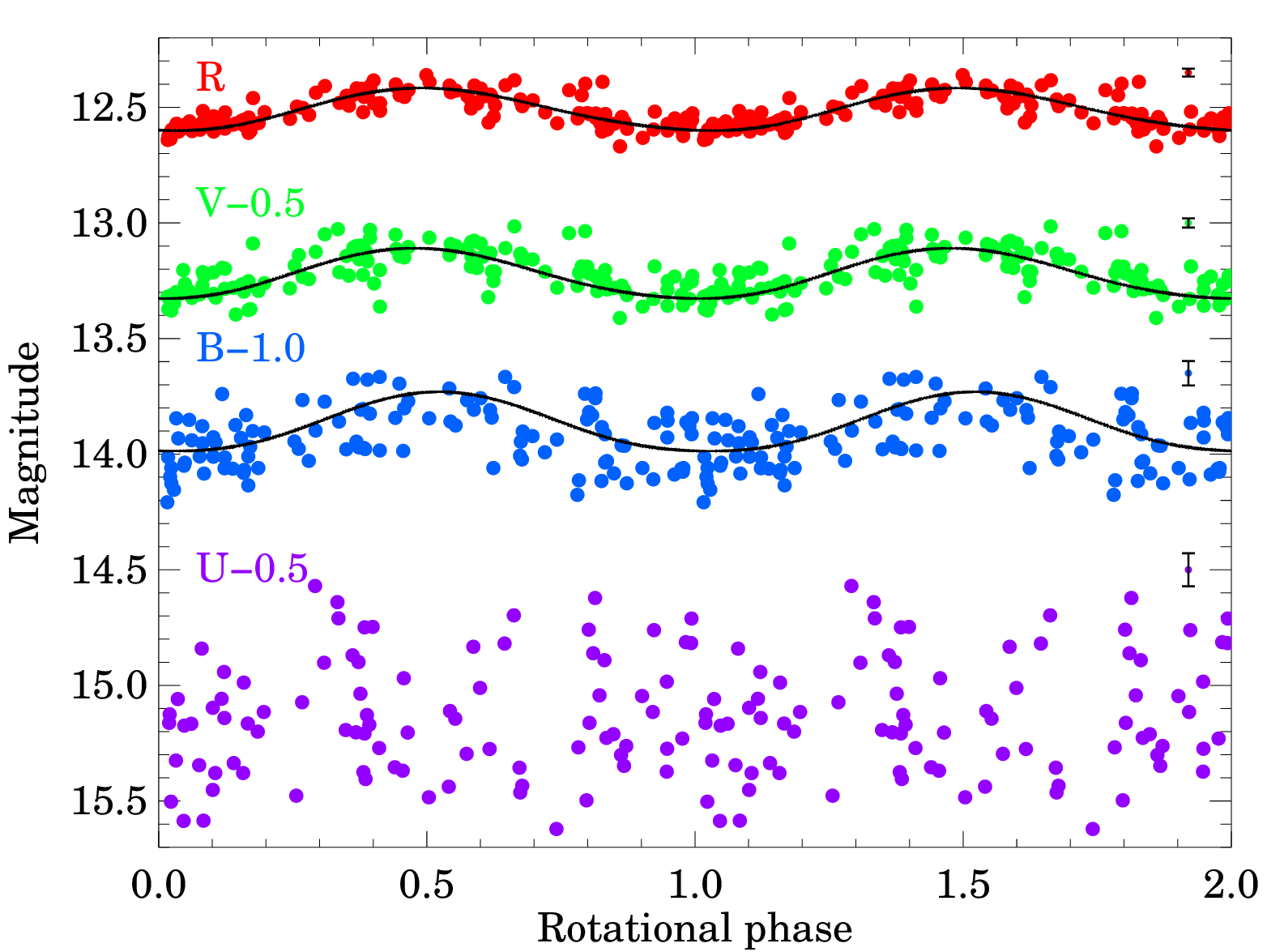}
  \includegraphics[angle=0,width=0.4\columnwidth]{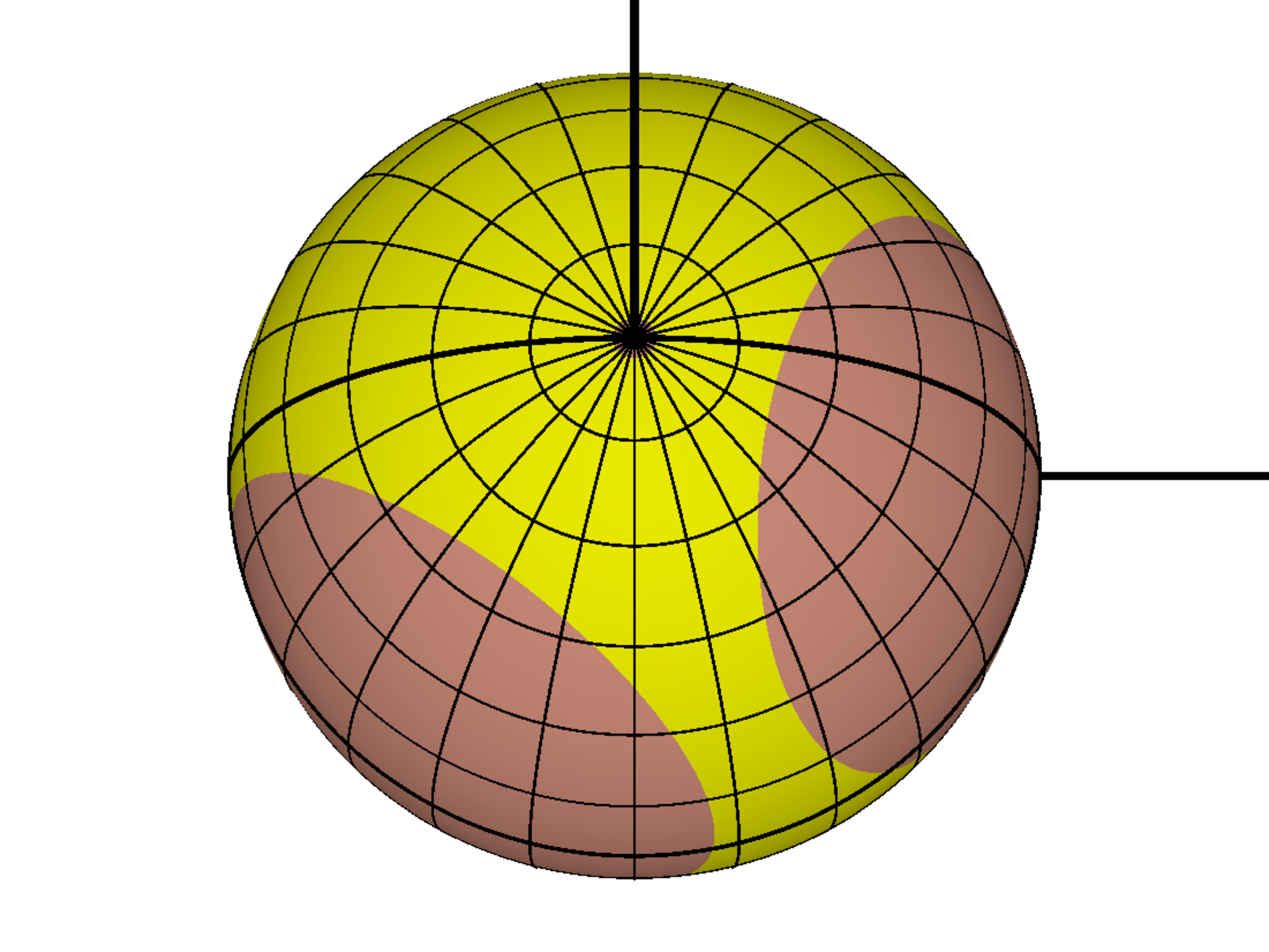}
  \caption{Top: phase-folded ground-based optical photometry of DQ~Tau
    based on data from \citet{tofflemire2017} and using $P$=3.017\,d,
    together with the brightness predicted by our fitted spot model
    (black curves). For clarity, the magnitudes in the different
    filters were shifted along the y axis by the values indicated in
    the graph. The typical error bar for each filter is displayed in
    the right-hand side. Bottom: a schematic representation of the
    spot distribution.
\label{fig:tofflemire_rotation}}
\end{center}
\end{figure}

We phase-folded \citet{tofflemire2017}'s data with the orbital period
of \citet{czekala2016} and discarded data points between phases
0.75--1.1, and also discarded points brighter than $U$=15.05\,mag,
$B$=14.65\,mag, $V$=13.50\,mag, and $R_C$=12.35\,mag, to exclude
possible flares and apastron accretion events. Then we phase-folded
the remaining points with a $P$=3.017\,d, the rotational period
obtained from the K2 light curve. The result, which can be seen in
Fig.~\ref{fig:tofflemire_rotation} is sinusoidal, at least in $BVR_C$,
because the $U$-band data is so noisy (or contaminated by flares) that
it might as well be consistent with constant brightness. We fitted a
sine curve to each of the $B$, $V$ and $R_C$ light curves keeping the
amplitude, magnitude offset, and phase as free parameters. The three
light curves gave the same phase.  As expected for stellar spots, the
variability amplitude we obtained is smaller towards longer
wavelengths, with peak-to-peak values of ${\Delta}B=0.20\,$mag,
${\Delta}V=0.20\,$mag, ${\Delta}R=0.17\,$mag.

As a next step, we tried to fit \citet{tofflemire2017}'s data with the
analytic spot model of \citet{budding1977}. We estimated the intensity
ratio between the quiet photosphere and the active regions by
blackbody models with appropriate temperatures. The linear limb
darkening coefficients were taken from \citet{claret2012}. A good fit
could be achieved using a model containing three circular spots with
homogeneous temperature. The stability of such models is discussed in
details in \citet{kovari1997}. By modeling the color indices, we found
that the temperature of the cool spots are $\approx$400\,K colder than
the stellar surface, for which we took 3600\,K (the average effective
temperature of the two components of the binary). The three spots are
located at longitudes of 230$^{\circ}$, 350$^{\circ}$, and
360$^{\circ}$ (with an uncertainty of $\pm$5$^{\circ}$), at latitudes
of 4$^{\circ}$, 25$^{\circ}$, and 90$^{\circ}$. These parameters were
fixed during the fitting, as there is only very limited information
encoded about these in the photometric data. The angular radii of the
spots are 43$^{\circ}$, 41$^{\circ}$, and 8$^{\circ}$ (with an
uncertainty of $\pm$5$^{\circ}$), meaning that the three spots
together cover about 50\% of the star. A polar spot is generally used
to account for the uncertainty in the unspotted brightness (the model
is sensitive to even very small changes of this parameter) and
long-term brightness variations, while the other two spots fit the
exact quasi-sinusoidal light curve shape. The synthetic light curves
produced by our spot model is overplotted in
Fig.~\ref{fig:tofflemire_rotation}. The bottom of the figure also
shows a schematic depiction of the spot coverage. Our model has
peak-to-peak variability amplitudes of ${\Delta}B=0.26\,$mag,
${\Delta}V=0.22\,$mag, ${\Delta}R=0.18\,$mag.

Our multi-filter $BV(RI)_C$ monitoring of DQ~Tau obtained during the
K2 campaign provided only 10 epochs, out of which 7 were taken close
to periastron. Only three are left to characterize the color of the
quiescent sinusoidal light variations. These data, within the
uncertainty, are consistent with the quiescent color variations
measured by \citet{tofflemire2017}. However, the 3.017\,d period
component of the K2 light curve displayed smaller variability
amplitude and a different phase. This suggests that during the K2
monitoring, the DQ~Tau system had a different spot coverage than
before, during \citet{tofflemire2017}'s monitoring. The similar colors
indicate that the temperature of the spots did not, while the
different phase and amplitude indicate that the location and extent of
the spots did change during the 2.5\,years that elapsed between their
observations and our K2 study. We obtained a phase shift of 0.73
compared to \citet{tofflemire2017}'s data, which corresponds to a
difference of about 97$^{\circ}$ (roughly a quarter rotation). We
could obtain a reasonably good fit by keeping the three spots in our
model at the same latitute as before, but shift their longitude and
decrease their radius to account for the smaller variability amplitude
(a likely result of surface evolution that can be also observed in the
quiet K2 data). On average, the spots covered about 30\% of the
stellar photosphere during the K2 observations. Moreover, we also
discovered that the spot coverage can change on a significantly
shorter timescale as well, as proven by the variable phases of the
sine functions during the 80 days of the K2 monitoring
(Fig.~\ref{fig:k2folded}).

It is an interesting question whether the spots are located on the
primary, on the secondary, or both stars are spotted. Our photometric
observations give no clue about this, and in our spot modeling, we
assumed only one star. In practice, considering that the two stars are
almost identical, having one star with about 50\% spot coverage and
another one with 0\%, or having two stars with 25\%-25\% spot coverage
would give the same light curves.  Therefore, the obtained results
should be used with caution. Detailed spectroscopic studies may reveal
fine details in the photospheric line profiles of each star, which may
help to decide the spot coverage of the individual components in the
future.

\subsection{Flares in the DQ~Tau system}

\begin{figure*}
  \includegraphics[angle=0,width=\textwidth]{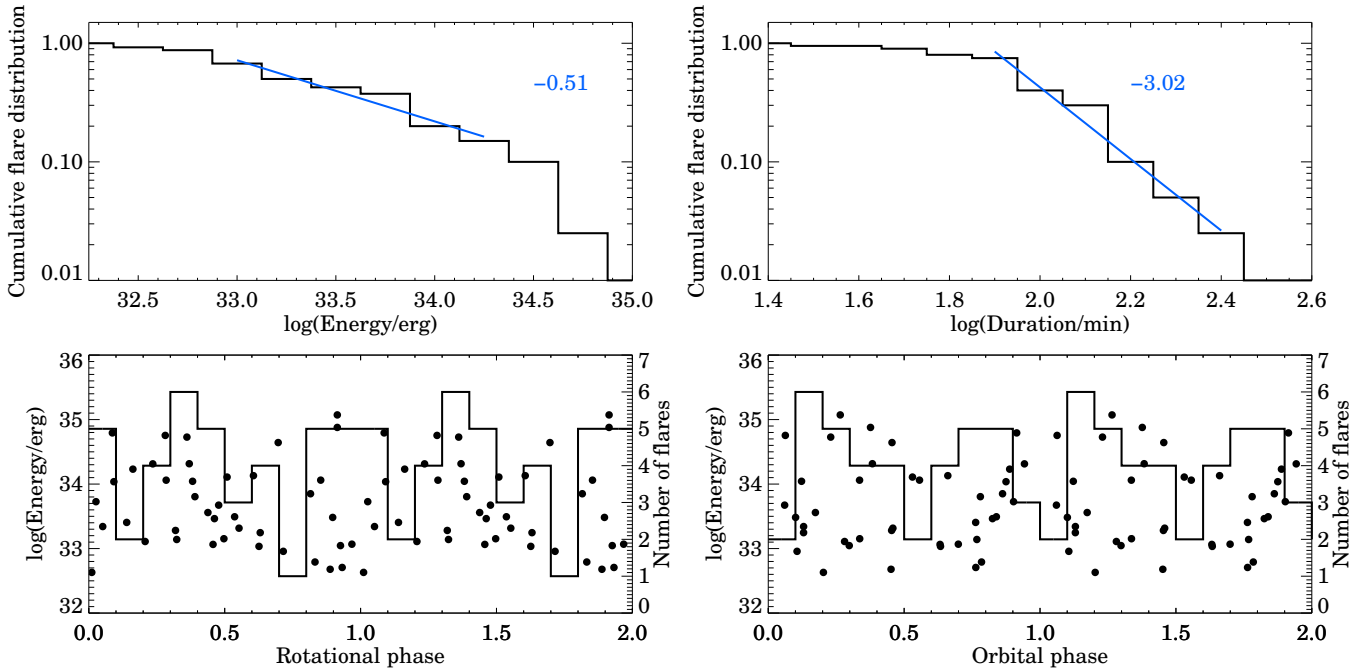}
  \caption{Flare characteristics in DQ~Tau.
\label{fig:histograms}}
\end{figure*}

By converting the K2 counts to power in erg\,s$^{-1}$ as described in
Section \ref{sec:obs}, and integrating the flare light curves over
time, we calculated the energy released in each flare. We obtained
values between 4.4$\times$10$^{32}$\,erg and
1.2$\times$10$^{35}$\,erg. The cumulative distribution of flare
energies is displayed in Fig.~\ref{fig:histograms}. Flares are often
described by a power-law trend of the energy--cumulative flare
frequency distribution \citep{gershberg1972}. If the number of flares
($dN$) in a given energy range ($E+dE$) is expressed as
$$dN(E) \propto E^{-\alpha} dE,$$ the cumulative flare distribution
should have a slope of $\beta = 1-\alpha$ in log-log scale (see also
\citealt{hawley2014} and \citealt{gizis2017}). For DQ~Tau,
Fig.~\ref{fig:histograms} indicates that apart from the lowest
energies (where completeness is not secure) and the highest energies
(which suffer from low number statistics), the cumulative distribution
can be fitted with a power law with an exponent of $\beta = -0.51$,
from which $\alpha = 1.51$. An alternative method to determine
$\alpha$ is to use the maximum likelihood estimator
\citep{arnold2014}:
$$
(\alpha - 1) = n \left[ \sum\limits_{i=1}^{n} \ln \frac{E_i}{E_{min}}\right] ^{-1},
$$ where $n$ is the number of flares detected, and $E_{min}$ and $E_i$
are the lowest and individual flare energies, respectively. For small
samples the result is biased, which can be corrected by multiplying by
a factor of $\frac{n-2}{n}$. This yields $\alpha=1.40$, which suggests
that the flares are of non-thermal origin
(cf.~\citealt{aschwanden2016}), and which places DQ~Tau close to the
very active M3--M5 sample of \citet{hawley2014}.

We also measured the duration of the flares as the time interval where
the points are higher than the 1$\sigma$ noise of the adjacent
portions of the light curves. The flares we found have durations
between 0.6 and 5.8 hours, with almost two thirds of the flares being
100--200 minutes long. The cumulative distribution of the flare
durations is quite steep, its power law exponent is $-$3.02
(Fig.~\ref{fig:histograms}).

By calculating for each flare the phase where they occurred, we
checked whether there is any rotational or orbital phase when flares
can preferentially be seen or are particularly energetic. We found,
however, that the corresponding (lower) panels of
Fig.~\ref{fig:histograms} display a seemingly random distribution. To
quantitatively decide whether the flares really occur at random
phases, we run one-sided Kolmogorov-Smirnov and Kuiper tests to assess
whether the phases are drawn from a uniform distribution. The obtained
probabilities were 0.90 and 0.81 for the orbital phases and 0.98 and
0.92 for the rotational phases, using the two different test,
respectively. This result means that the distribution of flares is not
significantly different from a random distribution. By applying
Wilcoxon rank test to compare the phase sample with a randomly
generated uniform distribution we came to the same conclusion. In
order to examine whether the energy of flares has any influence on the
phase distribution we divided the phases into two subsamples of equal
size that contain the 20 brightest and 20 faintest flares. Using
two-sided Kolmogorov-Smirnov and Kuiper tests, as well as the Wilcoxon
rank test, we found no statistically significant difference between
the two subsamples.

Concerning the orbital phases, this suggests that the flares are
probably not related to magnetospheric interactions, because their
occurrence is equally probable when the binary components are far away
from or close to each other, suggesting that they are more likely
single-star flares. Accepting the single-star scenario, the flares
could be happening in active regions above stellar spots, as seen on
the Sun. This could indicate a connection between the photospheric and
chromospheric activity, as reported for V374~Peg \citep{vida2016}. In
other stars, this connection is mainly shown as an anticorrelation of
the H$\alpha$ line intensity and the light curve intensity (e.g.,
EY~Dra, \citealt{korhonen2010}, on the T~Tauri-type star TWA~6,
\citealt{skelly2008}, but also on RS~CVn-type binaries like RT~Lac,
\citealt{frasca2002}). However, this connection is not always present:
on SAO~51891 no such relation was found \citep{biazzo2009}. If such a
connection holds for DQ~Tau, one would expect to observe more flares
around rotational phase 0, which corresponds to the faintest state of
the system, i.e., when the most spotted (and therefore most active)
side of the star faces us. While this is exactly what was observed in
LkCa~4 and LkCa~7 by \citet{vrba1993}, such effect is not evident from
our data on DQ~Tau. A possible reason for this is that such a large
fraction of the stellar surface is covered by spots that flares can be
observed practically at any time.

Because our K2 photometry is unresolved, there is no way to discern
whether the flares originated from the primary or the secondary. Since
the binary is composed of two almost identical stars, we can expect
similar flare activity from them both, and it would be reasonable to
assume that about half of the detected flares come from each
individual star. To account for this, we randomly removed half of the
flares and repeated our analyses described above. We found that our
main results do not change, therefore our main conclusions still hold.

We noted that for some flares (typically the brighter, single flares),
the flare template does not fit the light curves well. The residual
light curve from a fit to the peak and the first 5 minutes after the
peak shows a slow brightening followed by a long decrease of
intensity. This excess emission is unlike a typical flare signature,
therefore it must have a different origin, such as heating of one star
due to the energy released by a flare or coronal mass ejections (CMEs)
on the other star. We calculated the actual distance of the two stars
during flares \#1, \#21, and \#40 (0.07, 0.23 and 0.11\,au,
respectively), and the velocity needed to travel these distances, and
obtained values between $\approx 30\,000-100\,000$ km\,s$^{-1}$ for
five minute post-peak travel time. These values would vary only
slightly if we assume large magnetic loops of $\approx$1\,R$_*$. The
typical velocity of solar CMEs is on the order of a few hundred
km\,s$^{-1}$, and the fastest detected stellar CME (on the late-type
main sequence star AD~Leo) had a maximum projected velocity of
$\approx$5800\,km\,s$^{-1}$ \citep{houdebine1990}. These are an order
of magnitude smaller than the speed needed for the ejecta to reach the
other component in the DQ~Tau system, which makes the CME scenario
less likely. Direct heating by illumination, however, is still
possible.

\subsection{Pulsed accretion in the DQ~Tau system}

The accretion process from a circumbinary disk to the binary
components has been extensively studied in the literature by means of
numerical simulations. Although many of these works focus on
(super)massive black hole binary systems, some results are independent
of the applied distance or mass scales. For unequal-mass, eccentric
binaries, a common result of these models is that the accretion
happens in short bursts and displays a clear periodicity at the
binary's orbital period \citep{cuadra2009, roedig2011,
  sesana2012}. This is triggered by the secondary, which
gravitationally perturbs the inner edge of the circumbinary disk
during each apoapsis passage and pulls some material from the disk
that eventually lands on the binary components. Therefore, it is not
surprising that the secondary, which is closer to the disk and has a
greater specific angular momentum, accretes more mass
\citep{cuadra2009}. In some models, half and third of the orbital
period can also be seen in the periodogram of the accretion
rate. These are related to the second and third frequency harmonics
and are more pronounced for higher eccentricities
\citep{roedig2011}. Although DQ~Tau is closer to equal-mass than what
was used in these simulations ($M_1/M_2 = 0.33-0.35$), the orbital
period and its half appear in its periodogram.

In magnetohydrodynamic simulations of an equal-mass circular binary,
\citet{shi2012} found a gap radius (inner radius of the circumbinary
disk) of twice that of the binary separation and bursts in the
accretion rate at a period of about twice the binary's orbital
period. Hydrodynamic simulations by \citep{munoz2016} indicated that
equal-mass circular binaries display a quasi-periodic accretion rate
curve with a dominant period of 5 times the binary's orbital
period. This corresponds to the orbital period of the innermost region
of the circumbinary disk. On the other hand, the accretion rate in
eccentric binaries shows larger variability with pulses in the
accretion at a period equal to the binary's orbital period. Indeed, in
the almost equal-mass eccentric DQ~Tau system, we see large
variability in the accretion rate and a period close to the orbital
period. Interestingly, the gap radius in DQ~Tau is smaller and there
the Keplerian period coincides with the orbital period
(Sec.~\ref{sec:diskvariability}).

\citet{dorazio2013} made an extensive study of the accretion rate in
black hole binaries with different mass ratios. They also found that
the size of the inner gap is at most twice that of the binary
separation (for equal-mass binaries) and decreases for more extreme
mass ratios. An important result they obtained was that for binaries
between $q$=0.25--1 mass ratio, the accretion rate showed three
distinct periods, at 0.5, 1, and 5.7 times the binary's orbital
period. They claim that the last one is related to a dense lump of
material that formed in the inner disk created by shocks due to a
stream of material starting from the inner disk, and therefore depends
on the disk properties. However, the two periods with 1:2 ratio is a
robust prediction of their model, which is independent of the disk
properties, and could serve as an evidence for the presence of a
binary. Interestingly, the two most significant periods in DQ~Tau have
a ratio of 0.51$\pm$0.06, suggesting that the model predictions of
\citet{dorazio2013} for binary black holes may actually work for young
stellar binaries as well. This raises the possibility that if the
accretion-related variability in a young star's light curve displays
two significant periods with a 1:2 ratio, this may serve as a strong
hint for binarity. \citet{farris2014} went a step further and claimed
that the periodicities observed in the accretion rate may even be used
to infer the mass ratio of the binary.

\citet{gunther2002} modeled several circumbinary disks surrounding
classical T~Tau stars, including DQ~Tau. They found a flow of material
from the inner edge of the circumbinary disk onto the central stars
and that the accretion rate depends on the orbital phase. They could
match both the accretion rate and the spectral energy distribution of
the model to the observed properties of the DQ~Tau system. They found
minimal accretion rates of a few times
10$^{-9}\,M_{\odot}$\,yr$^{-1}$, and peak accretion rates between
1--2$\times$10$^{-8}\,M_{\odot}$\,yr$^{-1}$. According to their
simulations, the accretion rate both to the primary and to the
secondary peaked exactly at periastron. This is actually in contrast
to what we observe in DQ~Tau, where there are typically several peaks
per orbit and they all occur before the periastron, between phases of
0.8--1.0, see Fig.~\ref{fig:acc2}. The measured peak accretion rate
values agree well with the simulations. Another difference between our
accretion rate curve and that of \citet[][see their
  Fig.~12]{gunther2002} for DQ~Tau is that the simulated accretion
rate changes continuously, while in our observations the accretion
pulses are often separated by long quiescent periods when the
accretion rate is practically constant zero. This suggests a less
continuous mass flow and a lower average mass accretion rate than what
is obtained in the model.

\subsection{Disk variability in the DQ~Tau system}
\label{sec:diskvariability}

Our {\it Spitzer} 3.6 $\mu$m and 4.5 $\mu$m measurements are
distributed over a 11.4-day-long interval, which is 70\% of an orbital
period. The observations cover almost entirely the last periastron
event that occurred during the K2 monitoring, except the very steep
initial flux rise. There are also mid-infrared data points from the
quiescent part following the optical peak, and they cover the epoch of
the apastron, too.

The measured emission at the {\it Spitzer} wavelengths is a
combination of the stellar photosphere, modulated by the rotating
stellar spots, and the disk's thermal emission. None of our {\it
  Spitzer} epochs coincided with a stellar flare, thus, any possible
contribution from flares can be neglected in our case. In order to
determine the disk emission, first we subtracted from the measured
mid-infrared light curves a constant photospheric contribution
(0.187\,Jy and 0.098\,Jy at 3.6\,$\mu$m and 4.5\,$\mu$m, respectively,
based on our SED fitting explained in Section \ref{sec:obs}). Then we
used the spot model presented in Section \ref{sec:spotmodel}, and
calculated how the sinusoidal curve seen in the Kepler light curve
would appear at the {\it Spitzer} wavelengths using blackbody curves
for the stellar photosphere and for the spots as well. This periodic
signal was also subtracted from the {\it Spitzer} light curve, thus
the residual flux and its variability reflects the thermal emission of
the disk's dust particles only.

\begin{figure}
  \includegraphics[angle=90,width=\columnwidth]{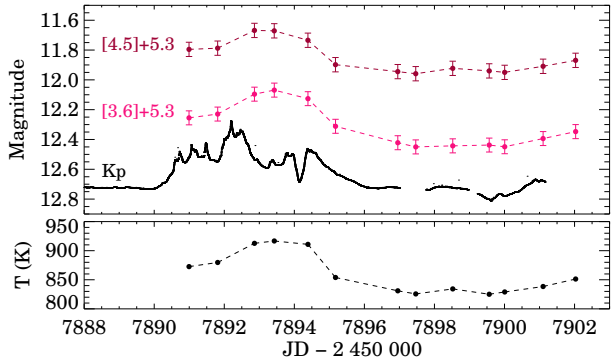}
  \caption{Top: K2 and {\it Spitzer} light curves of DQ~Tau after the
    removal of rotational modulation. Bottom: temperature variations
    in the inner disk of DQ~Tau based on blackbody fits to the {\it
      Spitzer} data.
\label{fig:spitzer}}
\end{figure}

The cleaned {\it Spitzer} light curve closely follows the Kepler
optical magnitude changes also cleaned from the photosphere, the spots
and the flare contributions (Fig.~\ref{fig:spitzer}). They both
outline a large periastron peak that lasts for about 5 days and hint
for a small local maximum at the very end of the K2 light curve, also
confirmed by the {\it Spitzer} data. This correlated behavior strongly
suggests that the variability in the disk's thermal emission is caused
by the changing optical brightness of the central stellar system.

In order to check for any possible time delay between the optical and
infrared data sets, which may provide information about the
irradiation process of the disk by the central stars, we performed a
cross-correlation analysis by interpolating the Kepler curve at the
{\it Spitzer} epochs, shifting the {\it Spitzer} light curve by
different time lags and calculating ${\chi}^2$ values between the two
data sets after transforming them to a common flux scale via a linear
transformation. Surprisingly, we obtained two equally deep minima in
the ${\chi}^2$ distribution. One constitutes a small delay of
${\Delta}t=1.44$\,h, the other corresponds to a shift of 10.32\,h. In
both cases the {\it Spitzer} light curve lagged behind the optical
one. Considering the relatively coarse sampling of the {\it Spitzer}
data set, the first minimum is consistent with no delay. The second
minimum with $10.32-1.44 = 8.88$\,h lag behind the first one may not
stem from a real phyical process, but may correspond to the average
time difference between neighboring localized narrow maxima during the
periastron event (Section \ref{sec:periastronbrightenings}). The
result that no significant delay was observed between the optical and
the infrared light curves suggests that the variable component of the
the mid-IR emission primarily originates from the optically thin
surface layer of the disk, where the temperature of the dust particles
can adapt to a variable radiation field within seconds
(cf.~\citealt{chiang1997}).

\begin{figure}
  \includegraphics[angle=0,width=\columnwidth]{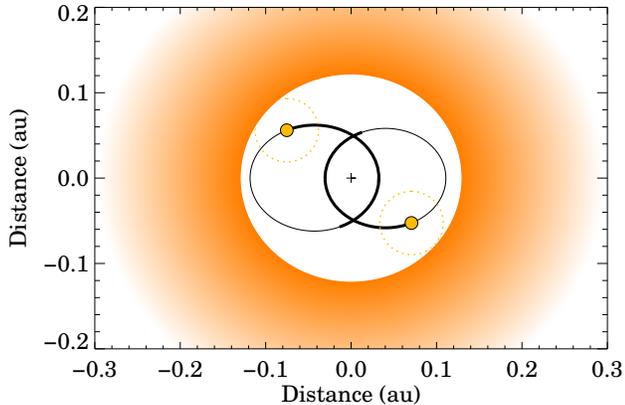}
  \caption{Sketch of the DQ Tau system. The + sign marks the center of
    mass. The black ellipses indicate the two stars' orbits using the
    orbital parameters and inclination from \citet{czekala2016}. The
    parts of the orbit when elevated accretion typically happens
    (between phases 0.75--1.1) are highlighted with thick black
    curves. The orange dots reflect the actual sizes of the stars
    using a stellar radius of 1.6$\,R_{\odot}$, while the dotted
    orange circles highlight the size of the magnetosphere for a
    typical T~Tauri star (5$\times$$R_*$, \citealt{hartmann1994}). The
    orange ring around the system indicates the inner part of the
    circumbinary disk, using 0.13\,au as its inner radius, see
    Sec.~\ref{sec:diskvariability} for details.
\label{fig:sketch}}
\end{figure}

The shapes and fluxes of the {\it Spitzer} light curves carry
information on the disk geometry. An estimate of the disk inner radius
can be derived from the {\it Spitzer} data. At each epoch, we fitted
the 3.6\,$\mu$m and 4.5\,$\mu$m fluxes (after subtracting the
photosphere and the spot components) by a single temperature blackbody
curve, assuming that the mid-infrared radiation is emitted from the
disk's inner edge. With a physically motivated prescription that the
main changes are related to the variable temperature while the
emitting area remains constant, we fixed the latter parameter at an
average value, and reproduced the {\it Spitzer} fluxes by varying the
temperature only. The obtained temperatures are plotted in the bottom
panel of Fig.~\ref{fig:spitzer}. We found the highest temperature
during the periastron event (917\,K), while the temperatures in the
quiescent interval were as low as 825\,K. Assuming equilibrium between
the impinging radiation field and the thermal dust emission in
quiescence, we can estimate the radius of the emitting region using
the following equation:
$$
\frac{L_*}{4\pi R^2} = \sigma T^4.
$$ From \citet{tofflemire2017} we adopted $L_* =
(0.19+0.13)\,L_{\odot} = 0.32\,L_{\odot}$ for the binary's luminosity
(for simplicity assuming that both stars are located at the center),
and obtained that the radius where the disk temperature is 825\,K is
$R = 0.13$\,au. We can consider this radius as the inner edge of the
(dust) disk (Fig.~\ref{fig:sketch}). Interestingly, this value is
equal to the binary's semimajor axis, while dynamical models of
circumbinary disks predict a cleared inner hole with a radius of
1.8--2.6 times larger that the semimajor axis
\citep{artymowicz1994}. Another interesting fact is that the Keplerian
orbital period at 0.13\,au is 15.56\,d, essentially identical to the
orbital period of the binary (and of course to the period of the
periastron brightenings). We can interpret this result as the inner
disk being in corotation with the orbit of the binary.

The absolute flux of the disk's thermal emission is determined by the
emitting area. Assuming that this area forms a ring whose inner radius
is $R_{\rm in}$=0.13\,au, we can calculate the outer radius as
well. If the emitting area is optically thick, and the disk's
inclination is 22$^{\circ}$ \citep{czekala2016}, almost pole-on, then
the outer radius of the emitting ring is $R_{\rm out} = 1.24\times
R_{\rm in}=0.16$\,au. We note that we adopted a uniform temperature
for this ring, but an outward decreasing radial temperature profile
would not change the conclusions significantly.

The ratio of the disk's luminosity (approximated by the integrated
flux of the 825\,K blackbody curve in quiescence) and the stellar
luminosity of 0.32$\,L_{\odot}$ is 0.163. We can interpret this number
as the solid angle of the inner disk as seen from the center of the
system (neglecting again the orbital motion of the stars). This means
that the height of the inner disk above the midplane is about $0.15
\times R_{\rm in}$, putting a constraint on the disk geometry.

While we computed a blackbody temperature of 825\,K for the quiescent
period from the {\it Spitzer} data, the temperature during the peak of
periastron event was higher, reaching 917\,K. If the emitting area was
the same as in quiescence, then the irradiating central luminosity had
to increase proportionally to $T^4$. This calculation shows that the
luminosity of the DQ Tau~system had to increase from quiescence to the
periastron peak by $(917/825)^4 = 1.53$, reaching
0.49$\,L_{\odot}$. Thus, the system emitted an excess luminosity of
$(0.49-0.32)\,L_{\odot}$ = 6.5$\times$10$^{32}$\,erg\,s$^{-1}$, which
is close to the luminosity peak measured in the {\it Kepler} band
during the last periastron event (taking into account the emission
outside the {\it Kepler} filter may bring the two numbers even closer
to each other). This result supports our scenario that the increasing
mid-infrared flux during the periastron event was due to increased
irradiation of the inner edge of the circumbinary disk at about
$0.13-0.16$\,au.

Finally we checked whether the free-fall time of the accreting
material is consistent with the computed inner disk radius. Again
placing both stars at the center for simplicity, we computed a
free-fall time of 2.7\,d from a distance of 0.13\,au. This value is
only half of the time difference between apastron (when the stars are
closest to the disk's inner edge and the perturbation of the disk is
expected) and the periastron (when the brightness peak implies the
infall of the material on the stellar surface). Considering the
complicated path the material may take from the inner edge of the disk
to the binary components, it is not at all unreasonable to assume that
this would take as much time as that elapsed between apastron and
periastron.

\subsection{Dipper phenomenon in DQ~Tau}
\label{sec:dipper}

Our results presented so far demonstrated the complexity of the
accretion process, as well as the circumstellar and magnetic structure
in the DQ~Tau binary system. After removing the spot-related
variability from the K2 light curve (Fig.~\ref{fig:corrected}), apart
from the flares and periastron events, another interesting phenomenon
can be seen: dips in the light curve when the system seems to be
fainter that what is expected from the stellar photosphere (also
taking into account the spots). These dips are marked by blue color in
Fig.~\ref{fig:corrected}. They are typically 0.3--1.6\,d long, and
their depth is between 0.04--0.09\,mag. They seem to be happening at
around rotational phase of 0.5, suggesting a possible connection with
the stellar rotation, although they are not strictly periodic. A
similar relationship was observed by \citet{bodman2017} using K2 data
on young stars in Upper Sco and $\rho$~Oph. We found no obvious
correlation with the orbital period.

The dipper phenomenon is explained as obscuration/eclipse by dusty
clumps of material close to the inner edge of the
disk. \citet{bodman2017} presented a unified paradigm of the dipper
phenomenon, and explained it by accretion streams that lift material
out of the disk midplane at the magnetospheric truncation radius. This
model also works for objects which are not completely edge-on (see
their Fig.~6), like DQ~Tau with its disk inclination of
22$^{\circ}$. Considering the complex magnetic structure of the
eccentric binary, we do not know where the magnetospheric trunction
radius is located in the DQ~Tau system. However, we know that the dust
disk's inner edge is probably around 0.13\,au (Section
\ref{sec:diskvariability}), where the Keplerian rotational period is
very close to the binary's orbital period. We speculate that the fact
that the disk's inner edge corotates with the binary may be due to a
common magnetic structure that the binary maintains. Thus, the
magnetic corotation radius in DQ~Tau may coincide with the dust disk's
inner edge. Therefore, the theory of \citet{bodman2017} is possibly
also applicable to DQ~Tau.

In this framework, we assume that the material causing the dips in
DQ~Tau's light curve is lifted up at the inner disk edge, at around
0.13\,au, then approaches the stars, and causes occultations despite
the almost pole-on orientation. We can estimate the mass of material
needed for the observed dips by calculating the column density from
the dips' depths using the relationship between the optical extinction
and hydrogen column density from \citet{guver2009}:
$$
N_H({\rm{cm}}^{-2}) = 2.22\times10^{21} A_V ({\rm{mag}}).
$$
Adopting 0.08\,mag as a typical dip depth in the {\it Kepler} band, we
can estimate a ${\Delta}A_V$ = 0.12\,mag inthe $V$ band using the
extinction curve of \citet{cardelli1989}. With this value, we obtained
a hydrogen column density of 2.6$\times$10$^{20}$\,cm$^{-2}$. If the
obscuring material passes in front of the stellar disk during 0.3\,d
with the Keplerian velocity at the disk's inner edge, and its width is
one stellar diameter, then its total mass is 1.4$\times$10$^{19}$\,g
or 7$\times$10$^{-15}\,M_{\odot}$, much less than the material
accreted in each orbital cycle (on the order of
10$^{-10}\,M_{\odot}$). Thus, it seems that a rather small amount of
mass is enough to explain the dips. Considering that the inner edge of
the disk is probably rather unstable due to the complex magnetic field
and highly time-dependent, dynamical perturbations by the binary
components as they orbit, it is not surprising to find such clumps of
material in the DQ~Tau system.

\subsection{Spin-orbit resonances}
\label{sec:resonance}

A spin-orbit resonance occurs when the rotational period and the
orbital period of a body have a simple integer ratio. Such resonances
can be found in a variety of objects such as binary asteroids
\citep{cuk2005}, planets in the Solar System \citep{colombo1965}, and
in some exoplanets as well \citep{szabo2012}. In order to better
understand the architecture of the DQ~Tau system, we performed an
analysis to look for possible spin-orbit resonances. This analysis is
enabled by the very precise rotational period derived here
($P=3.017\pm 0.004\,$d), and the orbital period precisely known from
radial velocity measurements ($P_{\rm orb}=15.80158\pm0.00066\,{\rm
  d}$ \citealt{tofflemire2017}).

\begin{figure}
  \includegraphics[angle=0,width=\columnwidth]{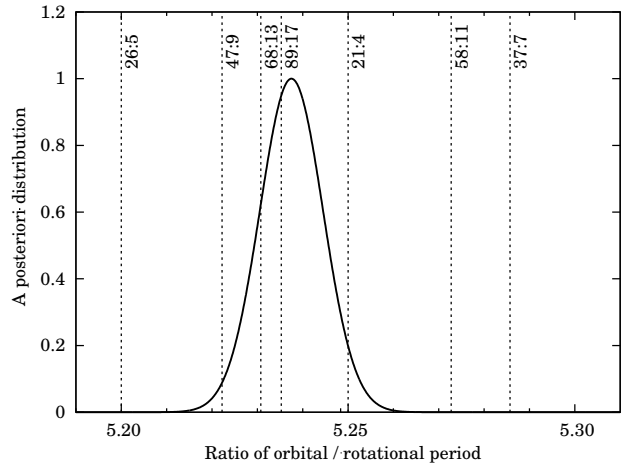}
  \caption{The \emph{a posteriori} distribution of the rotational
    period and the orbital period in the DQ~Tau system. Some of the
    possible resonances are marked with vertical dashed lines.
\label{fig:resonance}}
\end{figure}

We computed the ratio of the rotational period and the orbital period,
where the respective uncertainties were treated as uncorrelated and
Gaussian. The \emph{a posteriori} distribution of the ratio
$5.2375\pm0.0069$ is displayed in Fig.~\ref{fig:resonance}, along with
some of the possible resonances. The graph clearly shows that the peak
of the distribution avoids any low-order resonances.  Future
ground-based follow-up campaigns could aid to further constrain the
significance of a possible spin-orbit resonance.

%-----------------------------------------------------------------
% SUMMARY
%-----------------------------------------------------------------

\section{Summary}
\label{sec:sum}

We presented a detailed analysis of the light curves of the pre-main
sequence spectroscopic binary DQ~Tau, using very precise, very high
cadence 80-day-long optical photometric moitoring by K2, complemented
by multi-filter optical ground-based and mid-infrared {\it Spitzer}
photometry. We observed variability phenomena of various nature,
including rotational modulation by stellar spots, brief brightening
events due to stellar flares, long brigthening events around
periastron due to increased accretion, as well as short dips due to
brief circumstellar obscuration of the starlight. Our main findings
and conclusions are the following:

\begin{itemize}
\item DQ~Tau displays a complex light curve shape in which two main
  periods can be identified. One is a period of 3.017\,d corresponding
  to the stellar rotational period, which causes sinusoidal rotational
  modulation, while the other is a period of 15.6\,d due to powerful
  brightenings at each periastron of the binary. We found no
  spin-orbit resonance in the system.
\item By using our precise rotational period, we could identify
  similar rotational modulation in the $UBVRI$ monitoring data from
  \citet{tofflemire2017}, obtained 2.5\,yr before the K2
  monitoring. We fitted their photometry with a spot model containing
  a small polar cap, as well as two very extended spots together
  covering about 50\% of the stellar surface (assuming that the spots
  are on one of the binary components). The spots are 400\,K cooler
  than the stellar effective temperature of 3600\,K. Judging by the
  different variability amplitude and phase, the spot coverage and the
  location of the spots changed significantly during 2.5\,yr.
  Moreover, we detected signs of shifting phase and amplitude even
  within the 80-day-long K2 monitoring, suggesting rapidly changing
  spots.
\item After removing the rotational modulation from the K2 light
  curve, we could identify 40 flare events in the light curve,
  reminiscent of stellar flares often observed in late-type stars. The
  flares typically had a sharp rise and a slower decay, altogether
  lasting for 100--200\,minutes. The energy released in the flares was
  between 4.4$\times$10$^{32}$\,erg and 1.2$\times$10$^{35}$erg, more
  powerful than usual for main-sequence late-type stars, but typical
  for young low-mass stars. The energy distribution of the flares
  follows a rather shallow power law, suggesting non-thermal emission
  mechanism. The occurrence of the flares seems to be random; we saw
  no correlation either with the rotational period (probably a
  consequence of very extended active regions that are always
  visible), or with the binary's orbital period (suggesting that the
  flares are single-star flares happening just above the stellar
  surface rather than between the two stars where their magnetospheres
  periodically reconnect and separate). The most energetic flares, if
  they happen close to periastron, may result in the heating of one
  star by the flare of the other star.
\item After removing from the light curve the flares as well, the
  remaining brightness variations concentrate around the periastrons
  and are probably due to increased accretion from the circumbinary
  disk onto the stars. Using the K2 light curve as a proxy, we found
  that the peak accretion rates often reach
  5$\times$10$^{-8}$\,M$_{\odot}$\,yr$^{-1}$, there are typically
  three large peaks at each periastron, and the total accreted
  material is on average 1.2$\times$10$^{-10}$\,M$_{\odot}$ per
  orbital cycle.
\item Our {\it Spitzer} monitoring revealed that the increased
  accretion luminosity during a periastron event warmed up the inner
  disk by about 100\,K. By interpreting the region dominating the
  emission in the {\it Spitzer} bands as the inner edge of the dust
  disk, we found that it is at a distance of 0.13\,au, much less than
  expected for the cleared-out inner region around a
  binary. Interestingly, the inner edge is in corotation with the
  binary's orbit.
\item Finally, DQ~Tau also shows short dips of $<$0.1\,mag
  in its light curve, reminiscent of the well-known ``dipper
  phenomenon'' observed in many low-mass young stars. Similarly to the
  single stars, the explanations of the dips in the DQ~Tau system may
  be dusty material lifted up from the inner edge of the disk.
\end{itemize}

Our work on DQ~Tau demonstrated that high precision, high cadence
photometry together with simultaneous multi-filter data may be used to
disentangle the effects of rotation, flares, accretion, and
obscuration in young stellar objects. Although in the case of DQ~Tau
some of these phenomena are linked with the binarity of the target,
the applied methods are general and can be used for any young star
with the appropriate observations. In the context of binary physics,
our results may be used to verify numerical simulations of the
accretion process in young binaries, and may be even useful to
discover so far unknown binaries if they show similar light curve
features to those of DQ~Tau. It will also be a useful template to
follow when analyzing the light curves of known pre-main sequence
binaries (from, e.g., \citealt{mathieu1994} or \citealt{guenther2007})
provided by future missions like TESS or LSST.

%-----------------------------------------------------------------
% ACKNOWLEDGMENTS
%-----------------------------------------------------------------

\acknowledgments

This project has received funding from the European Research Council
(ERC) under the European Union’s Horizon 2020 research and innovation
programme under grant agreement No 716155 (SACCRED).  This work was
supported by the Momentum grant of the MTA CSFK Lend\"ulet Disk
Research Group. The authors acknowledge the Hungarian National
Research, Development and Innovation Office grants OTKA K-109276, OTKA
K-113117, OTKA K-115709, and supports through the Lend\"ulet-2012
Program (LP2012-31) of the Hungarian Academy of Sciences, and the ESA
PECS Contract No.~4000110889/14/NL/NDe. K.V.~is supported by the
Bolyai J\'anos Research Scholarship of the Hungarian Academy of
Sciences. This project has been supported by the
GINOP-2.3.2-15-2016-00003 grant of the Hungarian National Research,
Development and Innovation Office (NKFIH). This paper includes data
collected by the K2 mission. Funding for the K2 mission is provided by
the NASA Science Mission directorate. The authors wish to thank the
entire Kepler team and engineers for their persistence that made the
K2 Mission possible. This work is based in part on observations made
with the Spitzer Space Telescope, which is operated by the Jet
Propulsion Laboratory, California Institute of Technology under a
contract with NASA.

\facility{Kepler, Spitzer}.

%-----------------------------------------------------------------
% APPENDIX
%-----------------------------------------------------------------

\appendix

\begin{figure*}[h!]
  \includegraphics[angle=0,width=\textwidth]{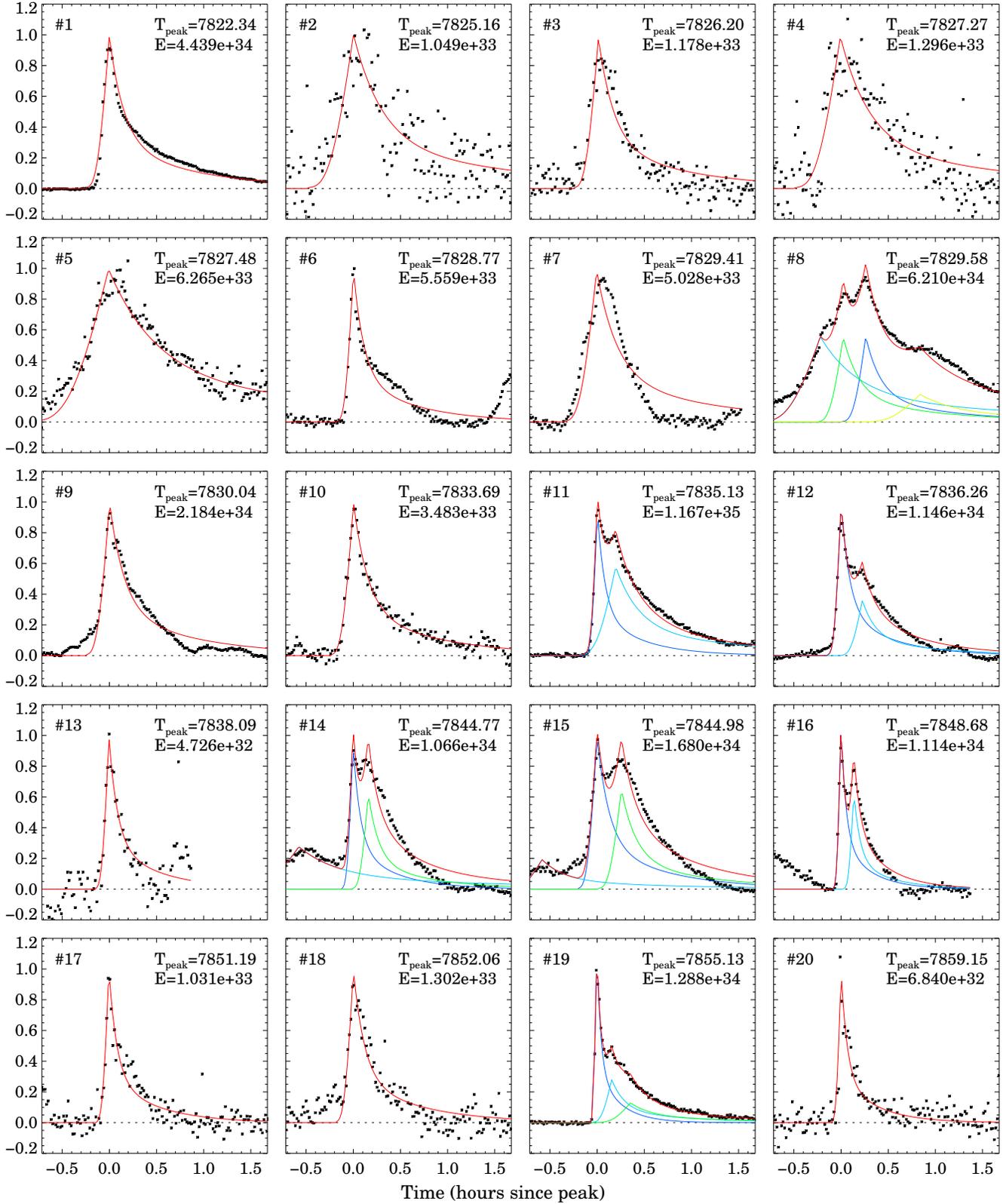}
  \caption{Flares in the K2 light curve of DQ~Tau. Counts are
    normalized so that the peak is 1. The numbers in the upper right
    corner of each panel indicate the JD$-$2\,450\,000 value of the
    peak of the flare and its total energy in ergs. The x axis range
    is 0.1\,d (2.4\,h or 144\,min) in all panels. Red curves display
    our fitted flare templates. In case of complex flares, the
    individual components are plotted with different colors, while the
    sum is indicated in red.
\label{fig:flares1}}
\end{figure*}

\begin{figure*}[h!]
  \includegraphics[angle=0,width=\textwidth]{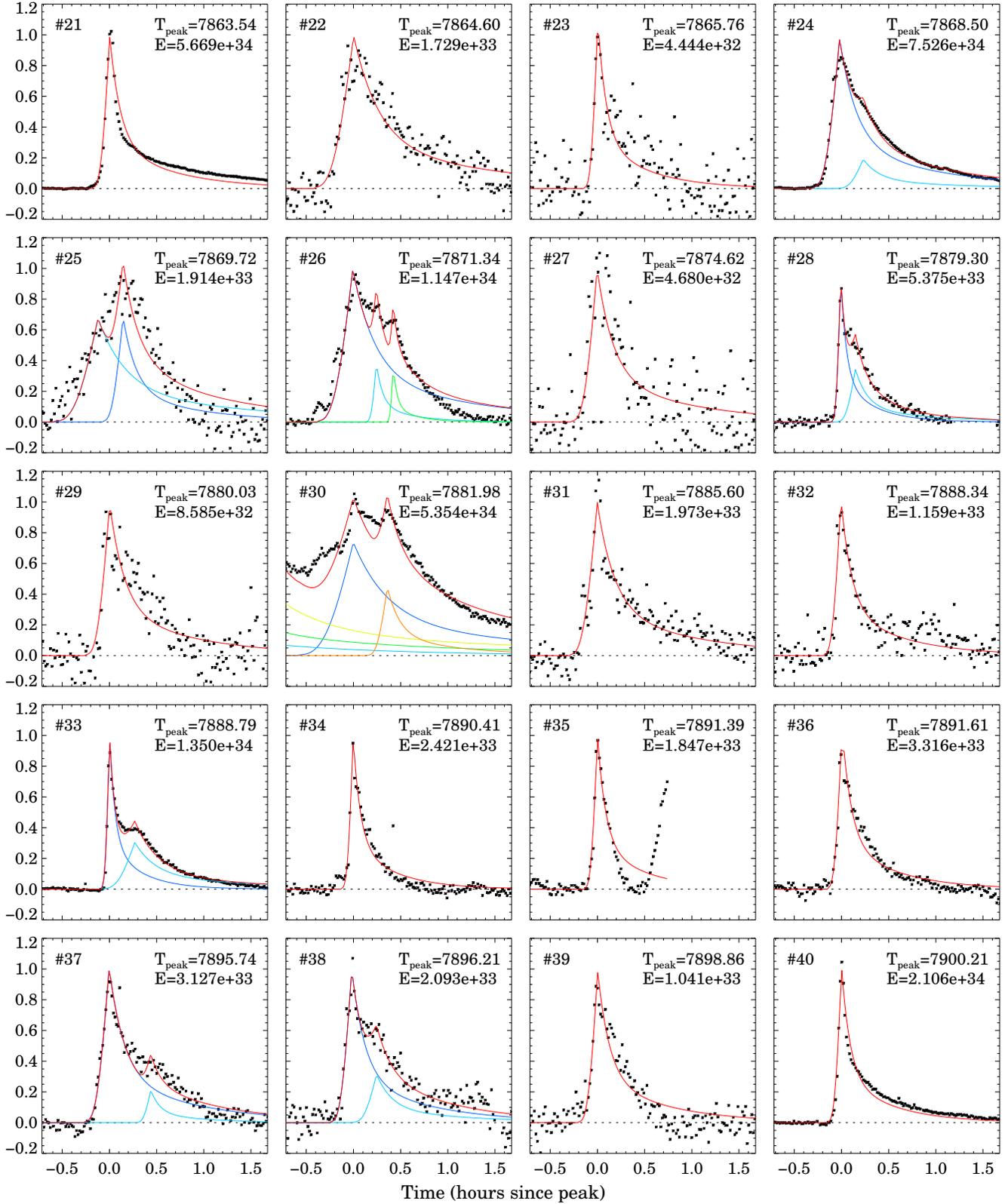}
  \caption{Flares in the K2 light curve of DQ~Tau (continued).
\label{fig:flares2}}
\end{figure*}

In Tab.~\ref{tab:photo} we present our ground-based optical and {\it
  Spitzer}-based infrared photometry for DQ~Tau. In
Figs.~\ref{fig:flares1} and \ref{fig:flares2} we show the light curves
of the flares identified in the K2 data on DQ~Tau.

\begin{table}[]
\centering
\caption{Ground-based optical and {\it Spitzer} mid-infrared photometry of DQ~Tau.}
\label{tab:photo}
\begin{tabular}{cccccccc}
\hline
\hline
Date & JD -- 2\,450\,000 & $B$  & $V$      & $R$      & $I$      & {[}3.6{]} & {[}4.5{]} \\
\hline
2017-03-08 & 7821.34 & 15.16(2) & 13.62(1) & 12.59(1) & 11.52(1) & \dots   & \dots   \\
2017-03-16 & 7829.27 & 14.43(1) & 13.22(1) & 12.27(1) & 11.25(1) & \dots   & \dots   \\
2017-03-28 & 7841.28 & 14.90(1) & 13.54(1) & 12.55(1) & 11.49(1) & \dots   & \dots   \\
2017-03-29 & 7842.27 & 14.76(1) & 13.45(1) & 12.41(1) & 11.41(2) & \dots   & \dots   \\
2017-03-30 & 7843.28 & 14.83(2) & 13.44(2) & 12.42(2) & 11.36(3) & \dots   & \dots   \\
2017-03-31 & 7844.29 & 14.30(1) & 13.17(1) & 12.22(1) & 11.25(1) & \dots   & \dots   \\
2017-04-02 & 7846.28 & 14.89(2) & 13.49(1) & 12.39(1) & 11.38(1) & \dots   & \dots   \\
2017-04-03 & 7847.28 & 14.73(2) & 13.46(1) & 12.41(1) & 11.39(3) & \dots   & \dots   \\
2017-04-09 & 7853.28 & 15.19(2) & 13.65(1) & 12.61(1) & 11.49(1) & \dots   & \dots   \\
2017-04-10 & 7854.28 & 15.12(2) & 13.62(2) & 12.57(1) & 11.47(1) & \dots   & \dots   \\
2017-05-17 & 7890.99 & \dots 	& \dots    & \dots    & \dots	 & 6.95(2) & 6.49(2) \\
2017-05-18 & 7891.81 & \dots	& \dots    & \dots    & \dots	 & 6.93(2) & 6.48(2) \\
2017-05-19 & 7892.86 & \dots	& \dots    & \dots    & \dots	 & 6.82(2) & 6.39(2) \\
2017-05-19 & 7893.43 & \dots	& \dots    & \dots    & \dots	 & 6.78(2) & 6.38(2) \\
2017-05-20 & 7894.39 & \dots	& \dots    & \dots    & \dots	 & 6.81(2) & 6.42(2) \\
2017-05-21 & 7895.18 & \dots	& \dots    & \dots    & \dots	 & 7.02(2) & 6.61(2) \\
2017-05-23 & 7896.95 & \dots	& \dots    & \dots    & \dots	 & 7.11(2) & 6.64(2) \\
2017-05-23 & 7897.47 & \dots	& \dots    & \dots    & \dots	 & 7.14(2) & 6.65(2) \\
2017-05-25 & 7898.52 & \dots	& \dots    & \dots    & \dots	 & 7.16(2) & 6.64(2) \\
2017-05-26 & 7899.56 & \dots	& \dots    & \dots    & \dots	 & 7.14(2) & 6.64(2) \\
2017-05-26 & 7900.01 & \dots	& \dots    & \dots    & \dots	 & 7.14(2) & 6.64(2) \\
2017-05-27 & 7901.10 & \dots	& \dots    & \dots    & \dots	 & 7.10(2) & 6.61(2) \\
2017-05-28 & 7902.03 & \dots	& \dots    & \dots    & \dots	 & 7.08(2) & 6.60(2) \\
\hline
\end{tabular}
\end{table}

%-----------------------------------------------------------------
% BIBLIOGRAPHY
%-----------------------------------------------------------------

\bibliography{paper}{}

\begin{thebibliography}{}
\expandafter\ifx\csname natexlab\endcsname\relax\def\natexlab#1{#1}\fi
\providecommand{\url}[1]{\href{#1}{#1}}

\bibitem[{Arnold(2014)}]{arnold2014}
Arnold, B.~C. 2014, Pareto Distribution (John Wiley \& Sons, Ltd).
\newblock \url{http://dx.doi.org/10.1002/9781118445112.stat01100.pub2}

\bibitem[{{Artymowicz} \& {Lubow}(1994)}]{artymowicz1994}
{Artymowicz}, P., \& {Lubow}, S.~H. 1994, \apj, 421, 651

\bibitem[{{Artymowicz} \& {Lubow}(1996)}]{artymowicz1996}
---. 1996, \apjl, 467, L77

\bibitem[{{Aschwanden} {et~al.}(2016){Aschwanden}, {Holman}, {O'Flannagain},
  {Caspi}, {McTiernan}, \& {Kontar}}]{aschwanden2016}
{Aschwanden}, M.~J., {Holman}, G., {O'Flannagain}, A., {et~al.} 2016, \apj,
  832, 27

\bibitem[{{Bailer-Jones} {et~al.}(2018){Bailer-Jones}, {Rybizki}, {Fouesneau},
  {Mantelet}, \& {Andrae}}]{bailerjones2018}
{Bailer-Jones}, C.~A.~L., {Rybizki}, J., {Fouesneau}, M., {Mantelet}, G., \&
  {Andrae}, R. 2018, ArXiv e-prints, arXiv:1804.10121

\bibitem[{{Basri} {et~al.}(1997){Basri}, {Johns-Krull}, \&
  {Mathieu}}]{basri1997}
{Basri}, G., {Johns-Krull}, C.~M., \& {Mathieu}, R.~D. 1997, \aj, 114, 781

\bibitem[{{Biazzo} {et~al.}(2009){Biazzo}, {Frasca}, {Marilli}, {Covino},
  {Alcal{\~a}}, {{\'C}akirli}, {Klutsch}, \& {Meyer}}]{biazzo2009}
{Biazzo}, K., {Frasca}, A., {Marilli}, E., {et~al.} 2009, \aap, 499, 579

\bibitem[{{Bodman} {et~al.}(2017){Bodman}, {Quillen}, {Ansdell}, {Hippke},
  {Boyajian}, {Mamajek}, {Blackman}, {Rizzuto}, \& {Kastner}}]{bodman2017}
{Bodman}, E.~H.~L., {Quillen}, A.~C., {Ansdell}, M., {et~al.} 2017, \mnras,
  470, 202

\bibitem[{{Bohlin} \& {Gilliland}(2004)}]{bohlin2004}
{Bohlin}, R.~C., \& {Gilliland}, R.~L. 2004, \aj, 127, 3508

\bibitem[{{Bouvier} {et~al.}(1999){Bouvier}, {Chelli}, {Allain}, {Carrasco},
  {Costero}, {Cruz-Gonzalez}, {Dougados}, {Fern{\'a}ndez}, {Mart{\'{\i}}n},
  {M{\'e}nard}, {Mennessier}, {Mujica}, {Recillas}, {Salas}, {Schmidt}, \&
  {Wichmann}}]{bouvier1999}
{Bouvier}, J., {Chelli}, A., {Allain}, S., {et~al.} 1999, \aap, 349, 619

\bibitem[{{Budding}(1977)}]{budding1977}
{Budding}, E. 1977, \apss, 48, 207

\bibitem[{{Cardelli} {et~al.}(1989){Cardelli}, {Clayton}, \&
  {Mathis}}]{cardelli1989}
{Cardelli}, J.~A., {Clayton}, G.~C., \& {Mathis}, J.~S. 1989, \apj, 345, 245

\bibitem[{{Castelli} \& {Kurucz}(2004)}]{castelli2004}
{Castelli}, F., \& {Kurucz}, R.~L. 2004, ArXiv Astrophysics e-prints,
  astro-ph/0405087

\bibitem[{{Chiang} \& {Goldreich}(1997)}]{chiang1997}
{Chiang}, E.~I., \& {Goldreich}, P. 1997, \apj, 490, 368

\bibitem[{{Claret} {et~al.}(2012){Claret}, {Hauschildt}, \&
  {Witte}}]{claret2012}
{Claret}, A., {Hauschildt}, P.~H., \& {Witte}, S. 2012, \aap, 546, A14

\bibitem[{{Cody} {et~al.}(2014){Cody}, {Stauffer}, {Baglin}, {Micela},
  {Rebull}, {Flaccomio}, {Morales-Calder{\'o}n}, {Aigrain}, {Bouvier},
  {Hillenbrand}, {Gutermuth}, {Song}, {Turner}, {Alencar}, {Zwintz},
  {Plavchan}, {Carpenter}, {Findeisen}, {Carey}, {Terebey}, {Hartmann},
  {Calvet}, {Teixeira}, {Vrba}, {Wolk}, {Covey}, {Poppenhaeger}, {G{\"u}nther},
  {Forbrich}, {Whitney}, {Affer}, {Herbst}, {Hora}, {Barrado}, {Holtzman},
  {Marchis}, {Wood}, {Medeiros Guimar{\~a}es}, {Lillo Box}, {Gillen},
  {McQuillan}, {Espaillat}, {Allen}, {D'Alessio}, \& {Favata}}]{cody2014}
{Cody}, A.~M., {Stauffer}, J., {Baglin}, A., {et~al.} 2014, \aj, 147, 82

\bibitem[{{Colombo}(1965)}]{colombo1965}
{Colombo}, G. 1965, \nat, 208, 575

\bibitem[{{Cuadra} {et~al.}(2009){Cuadra}, {Armitage}, {Alexander}, \&
  {Begelman}}]{cuadra2009}
{Cuadra}, J., {Armitage}, P.~J., {Alexander}, R.~D., \& {Begelman}, M.~C. 2009,
  \mnras, 393, 1423

\bibitem[{{{\'C}uk} \& {Burns}(2005)}]{cuk2005}
{{\'C}uk}, M., \& {Burns}, J.~A. 2005, \icarus, 176, 418

\bibitem[{{Czekala} {et~al.}(2016){Czekala}, {Andrews}, {Torres}, {Jensen},
  {Stassun}, {Wilner}, \& {Latham}}]{czekala2016}
{Czekala}, I., {Andrews}, S.~M., {Torres}, G., {et~al.} 2016, \apj, 818, 156

\bibitem[{{Davenport} {et~al.}(2012){Davenport}, {Becker}, {Kowalski},
  {Hawley}, {Schmidt}, {Hilton}, {Sesar}, \& {Cutri}}]{davenport2012}
{Davenport}, J.~R.~A., {Becker}, A.~C., {Kowalski}, A.~F., {et~al.} 2012, \apj,
  748, 58

\bibitem[{{Davenport} {et~al.}(2014){Davenport}, {Hawley}, {Hebb},
  {Wisniewski}, {Kowalski}, {Johnson}, {Malatesta}, {Peraza}, {Keil},
  {Silverberg}, {Jansen}, {Scheffler}, {Berdis}, {Larsen}, \&
  {Hilton}}]{davenport2014}
{Davenport}, J.~R.~A., {Hawley}, S.~L., {Hebb}, L., {et~al.} 2014, \apj, 797,
  122

\bibitem[{{Dodin}(2015)}]{dodin2015}
{Dodin}, A.~V. 2015, Astronomy Letters, 41, 196

\bibitem[{{D'Orazio} {et~al.}(2013){D'Orazio}, {Haiman}, \&
  {MacFadyen}}]{dorazio2013}
{D'Orazio}, D.~J., {Haiman}, Z., \& {MacFadyen}, A. 2013, \mnras, 436, 2997

\bibitem[{{Farris} {et~al.}(2014){Farris}, {Duffell}, {MacFadyen}, \&
  {Haiman}}]{farris2014}
{Farris}, B.~D., {Duffell}, P., {MacFadyen}, A.~I., \& {Haiman}, Z. 2014, \apj,
  783, 134

\bibitem[{{Favata} {et~al.}(2005){Favata}, {Flaccomio}, {Reale}, {Micela},
  {Sciortino}, {Shang}, {Stassun}, \& {Feigelson}}]{favata2005}
{Favata}, F., {Flaccomio}, E., {Reale}, F., {et~al.} 2005, \apjs, 160, 469

\bibitem[{{Feigelson} {et~al.}(2002){Feigelson}, {Garmire}, \&
  {Pravdo}}]{feigelson2002}
{Feigelson}, E.~D., {Garmire}, G.~P., \& {Pravdo}, S.~H. 2002, \apj, 572, 335

\bibitem[{{Fernandez} \& {Miranda}(1998)}]{fernandez1998}
{Fernandez}, M., \& {Miranda}, L.~F. 1998, \aap, 332, 629

\bibitem[{{Frasca} {et~al.}(2002){Frasca}, {{\c C}ak{\i}rl{\i}}, {Catalano},
  {Ibano{\v g}lu}, {Marilli}, {Evren}, \& {Ta{\c s}}}]{frasca2002}
{Frasca}, A., {{\c C}ak{\i}rl{\i}}, {\"O}., {Catalano}, S., {et~al.} 2002,
  \aap, 388, 298

\bibitem[{{Frasca} {et~al.}(2009){Frasca}, {Covino}, {Spezzi}, {Alcal{\'a}},
  {Marilli}, {F{\.z}r{\'e}sz}, \& {Gandolfi}}]{frasca2009}
{Frasca}, A., {Covino}, E., {Spezzi}, L., {et~al.} 2009, \aap, 508, 1313

\bibitem[{{Gershberg}(1972)}]{gershberg1972}
{Gershberg}, R.~E. 1972, \apss, 19, 75

\bibitem[{{Getman} {et~al.}(2016){Getman}, {Broos}, {K{\'o}sp{\'a}l}, {Salter},
  \& {Garmire}}]{getman2016}
{Getman}, K.~V., {Broos}, P.~S., {K{\'o}sp{\'a}l}, {\'A}., {Salter}, D.~M., \&
  {Garmire}, G.~P. 2016, \aj, 152, 188

\bibitem[{{Getman} {et~al.}(2011){Getman}, {Broos}, {Salter}, {Garmire}, \&
  {Hogerheijde}}]{getman2011}
{Getman}, K.~V., {Broos}, P.~S., {Salter}, D.~M., {Garmire}, G.~P., \&
  {Hogerheijde}, M.~R. 2011, \apj, 730, 6

\bibitem[{{Gizis} {et~al.}(2017){Gizis}, {Paudel}, {Mullan}, {Schmidt},
  {Burgasser}, \& {Williams}}]{gizis2017}
{Gizis}, J.~E., {Paudel}, R.~R., {Mullan}, D., {et~al.} 2017, \apj, 845, 33

\bibitem[{{Guenther} {et~al.}(2007){Guenther}, {Esposito}, {Mundt}, {Covino},
  {Alcal{\'a}}, {Cusano}, \& {Stecklum}}]{guenther2007}
{Guenther}, E.~W., {Esposito}, M., {Mundt}, R., {et~al.} 2007, \aap, 467, 1147

\bibitem[{{Gullbring}(1994)}]{gullbring1994}
{Gullbring}, E. 1994, \aap, 287, 131

\bibitem[{{Gullbring} {et~al.}(1998){Gullbring}, {Hartmann}, {Brice{\~n}o}, \&
  {Calvet}}]{gullbring1998}
{Gullbring}, E., {Hartmann}, L., {Brice{\~n}o}, C., \& {Calvet}, N. 1998, \apj,
  492, 323

\bibitem[{{G{\"u}nther} \& {Kley}(2002)}]{gunther2002}
{G{\"u}nther}, R., \& {Kley}, W. 2002, \aap, 387, 550

\bibitem[{{G{\"u}ver} \& {{\"O}zel}(2009)}]{guver2009}
{G{\"u}ver}, T., \& {{\"O}zel}, F. 2009, \mnras, 400, 2050

\bibitem[{{Hartmann} {et~al.}(1994){Hartmann}, {Hewett}, \&
  {Calvet}}]{hartmann1994}
{Hartmann}, L., {Hewett}, R., \& {Calvet}, N. 1994, \apj, 426, 669

\bibitem[{{Hawley} {et~al.}(2014){Hawley}, {Davenport}, {Kowalski},
  {Wisniewski}, {Hebb}, {Deitrick}, \& {Hilton}}]{hawley2014}
{Hawley}, S.~L., {Davenport}, J.~R.~A., {Kowalski}, A.~F., {et~al.} 2014, \apj,
  797, 121

\bibitem[{{Henden} {et~al.}(2015){Henden}, {Levine}, {Terrell}, \&
  {Welch}}]{apass}
{Henden}, A.~A., {Levine}, S., {Terrell}, D., \& {Welch}, D.~L. 2015, in
  American Astronomical Society Meeting Abstracts, Vol. 225, American
  Astronomical Society Meeting Abstracts \#225, 336.16

\bibitem[{{Herbst} {et~al.}(1994){Herbst}, {Herbst}, {Grossman}, \&
  {Weinstein}}]{herbst1994}
{Herbst}, W., {Herbst}, D.~K., {Grossman}, E.~J., \& {Weinstein}, D. 1994, \aj,
  108, 1906

\bibitem[{{Houdebine} {et~al.}(1990){Houdebine}, {Foing}, \&
  {Rodono}}]{houdebine1990}
{Houdebine}, E.~R., {Foing}, B.~H., \& {Rodono}, M. 1990, \aap, 238, 249

\bibitem[{{Jordi} {et~al.}(2006){Jordi}, {Grebel}, \& {Ammon}}]{jordi2006}
{Jordi}, K., {Grebel}, E.~K., \& {Ammon}, K. 2006, \aap, 460, 339

\bibitem[{{Kenyon} {et~al.}(1994){Kenyon}, {Dobrzycka}, \&
  {Hartmann}}]{kenyon1994}
{Kenyon}, S.~J., {Dobrzycka}, D., \& {Hartmann}, L. 1994, \aj, 108, 1872

\bibitem[{{K{\H{o}}v{\'a}ri} \& {Bartus}(1997)}]{kovari1997}
{K{\H{o}}v{\'a}ri}, Z., \& {Bartus}, J. 1997, \aap, 323, 801

\bibitem[{{Kochanek} {et~al.}(2017){Kochanek}, {Shappee}, {Stanek}, {Holoien},
  {Thompson}, {Prieto}, {Dong}, {Shields}, {Will}, {Britt}, {Perzanowski}, \&
  {Pojma{\'n}ski}}]{kochanek2017}
{Kochanek}, C.~S., {Shappee}, B.~J., {Stanek}, K.~Z., {et~al.} 2017, \pasp,
  129, 104502

\bibitem[{{Korhonen} {et~al.}(2010){Korhonen}, {Vida}, {Husarik}, {Mahajan},
  {Szczygie{\l}}, \& {Ol{\'a}h}}]{korhonen2010}
{Korhonen}, H., {Vida}, K., {Husarik}, M., {et~al.} 2010, Astronomische
  Nachrichten, 331, 772

\bibitem[{{Lehtinen} {et~al.}(2016){Lehtinen}, {Jetsu}, {Hackman}, {Kajatkari},
  \& {Henry}}]{lehtinen2016}
{Lehtinen}, J., {Jetsu}, L., {Hackman}, T., {Kajatkari}, P., \& {Henry}, G.~W.
  2016, \aap, 588, A38

\bibitem[{{Luger} {et~al.}(2016){Luger}, {Agol}, {Kruse}, {Barnes}, {Becker},
  {Foreman-Mackey}, \& {Deming}}]{luger2016}
{Luger}, R., {Agol}, E., {Kruse}, E., {et~al.} 2016, \aj, 152, 100

\bibitem[{{Markwardt}(2009)}]{markwardt2009}
{Markwardt}, C.~B. 2009, in Astronomical Society of the Pacific Conference
  Series, Vol. 411, Astronomical Data Analysis Software and Systems XVIII, ed.
  D.~A. {Bohlender}, D.~{Durand}, \& P.~{Dowler}, 251

\bibitem[{{Mathieu}(1994)}]{mathieu1994}
{Mathieu}, R.~D. 1994, \araa, 32, 465

\bibitem[{{Mathieu} {et~al.}(1997){Mathieu}, {Stassun}, {Basri}, {Jensen},
  {Johns-Krull}, {Valenti}, \& {Hartmann}}]{mathieu1997}
{Mathieu}, R.~D., {Stassun}, K., {Basri}, G., {et~al.} 1997, \aj, 113, 1841

\bibitem[{{Mu{\~n}oz} \& {Lai}(2016)}]{munoz2016}
{Mu{\~n}oz}, D.~J., \& {Lai}, D. 2016, \apj, 827, 43

\bibitem[{{Nguyen} {et~al.}(2012){Nguyen}, {Brandeker}, {van Kerkwijk}, \&
  {Jayawardhana}}]{nguyen2012}
{Nguyen}, D.~C., {Brandeker}, A., {van Kerkwijk}, M.~H., \& {Jayawardhana}, R.
  2012, \apj, 745, 119

\bibitem[{{Roedig} {et~al.}(2011){Roedig}, {Dotti}, {Sesana}, {Cuadra}, \&
  {Colpi}}]{roedig2011}
{Roedig}, C., {Dotti}, M., {Sesana}, A., {Cuadra}, J., \& {Colpi}, M. 2011,
  \mnras, 415, 3033

\bibitem[{{Romanova} {et~al.}(2008){Romanova}, {Kulkarni}, \&
  {Lovelace}}]{romanova2008}
{Romanova}, M.~M., {Kulkarni}, A.~K., \& {Lovelace}, R.~V.~E. 2008, \apjl, 673,
  L171

\bibitem[{{Rydgren} \& {Vrba}(1983)}]{rydgren1983}
{Rydgren}, A.~E., \& {Vrba}, F.~J. 1983, \apj, 267, 191

\bibitem[{{Salter} {et~al.}(2008){Salter}, {Hogerheijde}, \&
  {Blake}}]{salter2008}
{Salter}, D.~M., {Hogerheijde}, M.~R., \& {Blake}, G.~A. 2008, \aap, 492, L21

\bibitem[{{Salter} {et~al.}(2010){Salter}, {K{\'o}sp{\'a}l}, {Getman},
  {Hogerheijde}, {van Kempen}, {Carpenter}, {Blake}, \& {Wilner}}]{salter2010}
{Salter}, D.~M., {K{\'o}sp{\'a}l}, {\'A}., {Getman}, K.~V., {et~al.} 2010,
  \aap, 521, A32

\bibitem[{{Sesana} {et~al.}(2012){Sesana}, {Roedig}, {Reynolds}, \&
  {Dotti}}]{sesana2012}
{Sesana}, A., {Roedig}, C., {Reynolds}, M.~T., \& {Dotti}, M. 2012, \mnras,
  420, 860

\bibitem[{{Shappee} {et~al.}(2014){Shappee}, {Prieto}, {Grupe}, {Kochanek},
  {Stanek}, {De Rosa}, {Mathur}, {Zu}, {Peterson}, {Pogge}, {Komossa}, {Im},
  {Jencson}, {Holoien}, {Basu}, {Beacom}, {Szczygie{\l}}, {Brimacombe},
  {Adams}, {Campillay}, {Choi}, {Contreras}, {Dietrich}, {Dubberley},
  {Elphick}, {Foale}, {Giustini}, {Gonzalez}, {Hawkins}, {Howell}, {Hsiao},
  {Koss}, {Leighly}, {Morrell}, {Mudd}, {Mullins}, {Nugent}, {Parrent},
  {Phillips}, {Pojmanski}, {Rosing}, {Ross}, {Sand}, {Terndrup}, {Valenti},
  {Walker}, \& {Yoon}}]{shappee2014}
{Shappee}, B.~J., {Prieto}, J.~L., {Grupe}, D., {et~al.} 2014, \apj, 788, 48

\bibitem[{{Shi} {et~al.}(2012){Shi}, {Krolik}, {Lubow}, \& {Hawley}}]{shi2012}
{Shi}, J.-M., {Krolik}, J.~H., {Lubow}, S.~H., \& {Hawley}, J.~F. 2012, \apj,
  749, 118

\bibitem[{{Skelly} {et~al.}(2008){Skelly}, {Unruh}, {Collier Cameron},
  {Barnes}, {Donati}, {Lawson}, \& {Carter}}]{skelly2008}
{Skelly}, M.~B., {Unruh}, Y.~C., {Collier Cameron}, A., {et~al.} 2008, \mnras,
  385, 708

\bibitem[{{Stassun} \& {Wood}(1999)}]{stassun1999}
{Stassun}, K., \& {Wood}, K. 1999, \apj, 510, 892

\bibitem[{{Stassun} {et~al.}(2006){Stassun}, {van den Berg}, {Feigelson}, \&
  {Flaccomio}}]{stassun2006}
{Stassun}, K.~G., {van den Berg}, M., {Feigelson}, E., \& {Flaccomio}, E. 2006,
  \apj, 649, 914

\bibitem[{{Stelzer} {et~al.}(2000){Stelzer}, {Neuh{\"a}user}, \&
  {Hambaryan}}]{stelzer2000}
{Stelzer}, B., {Neuh{\"a}user}, R., \& {Hambaryan}, V. 2000, \aap, 356, 949

\bibitem[{{Still} \& {Barclay}(2012)}]{still2012}
{Still}, M., \& {Barclay}, T. 2012, {PyKE: Reduction and analysis of Kepler
  Simple Aperture Photometry data}, Astrophysics Source Code Library, , ,
  ascl:1208.004

\bibitem[{{Szab{\'o}} {et~al.}(2012){Szab{\'o}}, {P{\'a}l}, {Derekas}, {Simon},
  {Szalai}, \& {Kiss}}]{szabo2012}
{Szab{\'o}}, G.~M., {P{\'a}l}, A., {Derekas}, A., {et~al.} 2012, \mnras, 421,
  L122

\bibitem[{{Tofflemire} {et~al.}(2017){Tofflemire}, {Mathieu}, {Ardila},
  {Akeson}, {Ciardi}, {Johns-Krull}, {Herczeg}, \&
  {Quijano-Vodniza}}]{tofflemire2017}
{Tofflemire}, B.~M., {Mathieu}, R.~D., {Ardila}, D.~R., {et~al.} 2017, \apj,
  835, 8

\bibitem[{{Vanderburg} \& {Johnson}(2014)}]{vanderburg2014}
{Vanderburg}, A., \& {Johnson}, J.~A. 2014, \pasp, 126, 948

\bibitem[{{Vida} {et~al.}(2016){Vida}, {Kriskovics}, {Ol{\'a}h}, {Leitzinger},
  {Odert}, {K{\H o}v{\'a}ri}, {Korhonen}, {Greimel}, {Robb}, {Cs{\'a}k}, \&
  {Kov{\'a}cs}}]{vida2016}
{Vida}, K., {Kriskovics}, L., {Ol{\'a}h}, K., {et~al.} 2016, \aap, 590, A11

\bibitem[{{Vrba} {et~al.}(1993){Vrba}, {Chugainov}, {Weaver}, \&
  {Stauffer}}]{vrba1993}
{Vrba}, F.~J., {Chugainov}, P.~F., {Weaver}, W.~B., \& {Stauffer}, J.~S. 1993,
  \aj, 106, 1608

\bibitem[{{Vrba} {et~al.}(1988){Vrba}, {Herbst}, \& {Booth}}]{vrba1988}
{Vrba}, F.~J., {Herbst}, W., \& {Booth}, J.~F. 1988, \aj, 96, 1032

\bibitem[{{Wood} {et~al.}(1996){Wood}, {Kenyon}, {Whitney}, \&
  {Bjorkman}}]{wood1996}
{Wood}, K., {Kenyon}, S.~J., {Whitney}, B.~A., \& {Bjorkman}, J.~E. 1996,
  \apjl, 458, L79

\end{thebibliography}

%-----------------------------------------------------------------
% THE END
%-----------------------------------------------------------------

\end{document}